\documentclass{article}
\usepackage{amssymb}

\usepackage{graphicx}
\usepackage{amsmath}

\newtheorem{theorem}{Theorem}
\newtheorem{acknowledgement}[theorem]{Acknowledgement}

\newtheorem{definition}[theorem]{Definition}

\newtheorem{remark}[theorem]{Remark}

\input{tcilatex}

\begin{document}

\title{New Concepts in Particle Physics from Solution of an Old Problem}
\author{Bert Schroer \\
Institut f\"{u}r Theoretische Physik\\
FU-Berlin, Arnimallee 14, 14195 Berlin, Germany\\
presently: CBPF, Rua Dr. Xavier Sigaud, 22290-180 Rio de Janeiro, Brazil \\
schroer@cbpf.br}
\date{July 1999, revised March 2000, to be published in JPA}
\maketitle

\begin{abstract}
Recent ideas on modular localization in local quantum physics are used to
clarify the relation between on- and off-shell quantities in particle
physics; in particular the relation between on-shell crossing symmetry and
off-shell Einstein causality. Among the collateral results of this new
nonperturbative approach are profound relations between crossing symmetry of
particle physics and Hawking-Unruh like thermal aspects (KMS property,
entropy attached to horizons) of quantum matter behind causal horizons,
aspects which hitherto were exclusively related with Killing horizons in
curved spacetime rather than with localization aspects in Minkowski space
particle physics. The scope of this modular framework is amazingly wide and
ranges from providing a conceptual basis for the d=1+1 bootstrap-formfactor
program for factorizable d=1+1 models to a decomposition theory of QFT's in
terms of a finite collection of unitarily equivalent chiral conformal
theories placed a specified relative position within a common Hilbert space
(in d=1+1 a holographic relation and in higher dimensions more like a
scanning). The new framework gives a spacetime interpretation to the
Zamolodchikov algebra and explains its thermal aspects.
\end{abstract}

\section{Introduction}

Theoretical physicists, contrary to mathematicians, rarely return to their
old unsolved problems; often they replace them by new inventions. The
content of the present article on some new concepts in particle physics does
not follow this pattern. The old problems it addresses and partially solves
are those of the relation between off-shell and on-shell quantities (or
between fields-particles) and in particular of crossing symmetry in local
quantum physics (LQP)\footnote{%
We will often use the name ''local quantum physics'' (LQP) instead of QFT 
\cite{Ha}, if we have in mind the physical principles of QFT implemented by
different concepts than those of the various quantization formalisms
(canonical, quantization via path integrals etc.) which most of the readers
are familiar with from the various textbooks. To the extend that the reader
does not automatically identify QFT with those formalisms, he may without
danger of misunderstandings continue to use the good old name QFT.}. A more
restricted form of on-shell crossing symmetry also led to the invention of
the dual model and string theory, a line of development which we will not
follow except some remarks in the last section.

The most prominent of on-shell quantities is the S-matrix of a local QFT,
whereas fields and more general operators are ``off-shell''. The derivation
of on-shell quantities from LQP through the use of the rigorous LSZ
scattering theory was one of the high points of the QFT of the 60$^{ies}.$
In the opposite direction the problem (``\textit{The} Inverse Problem of
QFT'') lay dormant for a long time. Recently the adaptation of the
Tomita-Takesaki modular theory to wedge-localized algebras suggested new
methods to construct unique off-shell local operator algebras from the
scattering data in a quite interesting and novel way \cite{Sch1}\cite{S-W2}.
Thus the inverse problem of QFT has a much better status than in QM. To
bring this into the open requires the introduction of a wealth of new
concepts relevant to particle physics, while maintaining all the principles
of QFT.

In this paper we will have to study a new kind of operators which, as a
result of their weak semiinfinite (wedge-like) localization and their close
relation to the S-matrix, are to be considered as on-shell. This on-shell
operators are essential for our new approach which avoids pointlike fields
at the beginning and rather starts with on-shell generators of
wedge-localized algebra which encode the structure of the S-matrix.
Off-shell \textit{compactly} localized operators and local field generators
are then obtained via intersections of wedge algebras. Here and in the
sequel the word localization region always stand for the causal completion
of a spacetime region; these are typically the regions which one obtains by
intersecting wedges.

Besides these two extremes there are intermediate possibilities where
on-shell and off-shell aspects appear together. The most prominent and
useful mixed objects are bilinear forms on scattering vector states i.e.
matrix elements of local operators $A$ (either pointlike fields or bounded
operators localized in smaller than wedge regions) taken between incoming
and outgoing multiparticle scattering states (in terms of Feynman graphs,
one leg is off-shell). 
\begin{equation}
^{out}\left\langle q_{1},...q_{n-1},q_{n}\left| A\right|
p_{n},p_{n-1},...p_{1}\right\rangle ^{in}  \label{1}
\end{equation}
which we will call (generalized) formfactors, following the standard
terminology of d=1+1 factorizing models. These objects fulfil the important
crossing symmetry which acts on the on-shell momenta.

\begin{eqnarray}
&&^{out}\left\langle q_{1},...q_{n-1},q_{n}\left| A\right|
p_{n},p_{n-1},...p_{1}\right\rangle ^{in} \\
&=&^{out}\left\langle -\bar{p}_{1,}q_{1},...q_{n-1},q_{n}\left| A\right|
p_{n},p_{n-1},...p_{2}\right\rangle ^{in}  \notag
\end{eqnarray}
where the analytic continuation $p\rightarrow -p$ is carried out in the
rapidity parametrization by an $i\pi $-shift: $\theta \rightarrow \theta
+i\pi ,$ and the bar denotes the antiparticle. The difficulties in physical
interpretation of this relation (about which rigorous information outside of
perturbation theory of sufficient generality is scarce) reflects the lack of
its conceptual understanding. It is in a way deeper than the TCP-symmetry, a
symmetry derived from causality which among other things requires the
existence of an antiparticle to each particle. In fact the crossing
transformation is a kind of individual TCP-transformation which effects only
one particle at a time\ within the multiparticle incoming ket configuration
and carries it to the outgoing bra configuration as an antiparticle. In
spite of its name it is not a quantum theoretical (Wigner) symmetry, since
that crossing process involves an on-shell analytic continuation $%
p\rightarrow -p$. Together with vacuum polarization it belongs to the most
characteristic aspects of QFT. Although its physical meaning in terms of the
basic principles of LQP remained vague, most physicists liked to view it as
a kind of on-shell imprint of Einstein causality, the latter being an
off-shell concept. One of the results of the new conceptual framework
presented here is an interpretation of the crossing property in terms of \
``wedge localization'' and the ensuing thermal Hawking-Unruh properties.
They are usually associated exclusively with black hole quantum physics, but
in fact turn out to be general properties of any local quantum description
including particle physics in Minkowski space. Although in constructive
terms the control in passing from the on-shell S-matrix-dominated aspects to
the off-shell local quantum physics remains a formidable problem, it is easy
to see that a local theory is (if it exists at all) uniquely determined in
terms of its on-shell ``shadow''.

The S-matrix whose matrix elements result from the previous formula for $A$ $%
=1,$ is \textit{the} observable of particle physics par excellence; it is
totally intrinsic and independent of any field coordinatizations, although
in the LSZ theory it is calculated from specific fields. Strictly speaking
in high energy physics only (inclusive) cross sections and not amplitudes
are directly measured; a fact which is especially important if interactions
between zero mass particles leads to infrared problems.

The reason why most theoretical methods in particle physics do not aim
directly at the S-matrix is that most of our physical intuition about
causality and charge flows in spacetime is based on (off-shell) local fields
or local observables. The new on-shell wedge-algebra generators introduced
in this paper are somewhat hidden and in particular are not obtainable by
Lagrangian or more generally by any kind of quantization approach\footnote{%
Any approach which leads to explicite solution and not just to formal
representations as e.g. euclidean functional integral representations would
of cause present all properties. But the Lagrangian approach only achieves
the latter.}. Although their role in general QFT is only at the beginning of
being understood, there is already a very good spacetime comprehension in
the class of factorizing d=1+1 models \cite{Sch1}.

The old problems on which there has been significant recent progress can be
compressed in terms of the following questions

\begin{itemize}
\item  Does a physically admissable S-matrix fulfilling unitarity, crossing
symmetry and certain analytic properties (needed in its formulation), have
an underlying unique local QFT? This one may call the \textit{inverse
problem of QFT associated with scattering. }It is a problem of principle
interest to take notice not only of the well-known fact that fields and
local observable lead to scattering, but that also local equivalence classes
of fields\footnote{%
That an S-matrix cannot determine individual fields had been known since the
late 50$^{ies}.$} or nets of local observables are in turn determined by
particle scattering data.

\item  Is there a constructive procedure in which, similar to the d=1+1
bootstrap-formfactor program for factorizing d=1+1 models (which in fact
reappears as a special case), the S-matrix and the generalized formfactors
enter as important constructive elements in order to obtain off-shell
objects as fields or local observables? In particular can one formulate such
a constructive approach in a conceptually intrinsic manner i.e. without any
quantization parallelism to classical field theory and without the use of
field coordinatizations and short-distance divergence problems? This could
be of tremendous practical importance.
\end{itemize}

The progress obtained on both questions by the modular method be presented
in the sequel.

In order to understand better what is meant by the word ''old'' in the title
of this paper, it is very instructive to pause and take stock of some past
achievements and failures in an S-matrix approach to particle physics.
Already as far back as 1946 Heisenberg \cite{Hei} proposed to do particle
physics in a pure S-matrix setting in order to avoid the at that time
nonsensical aspects of short distance divergencies in QFT. His requirements
of unitarity, Poincar\'{e} invariance and some aspects of cluster
decomposition properties turned out to be much too general in order to be
useful. A second attempt with the full backing of renormalized perturbation
theory was launched in the early 60$^{ies}$ \cite{Chew}$.$ Part of the
motivation was similar to the previous one. Although renormalization theory
meanwhile allowed to extract a class of perturbatively finite QFTs, the
formally infinite intermediate steps and the not entirely natural but rather
technical looking division into renormalizable/nonrenormalizable models
nourished the hope that those discomforting features would disappear in a
pure S-matrix approach \cite{White}. The second more pragmatic motivation
was the idea that the dispersion theoretical research of the 50$^{ies}$
could be extended into a computational scheme for strong interactions within
an S-matrix setting. Since the physical principles of the S-matrix approach
as any physical principles cannot be dealt with directly but rather require
an operator or functional formalism, several ideas which could not be
motivated through QFT (in fact they are not true in perturbative QFT) had to
be added. The most prominent ones were on-shell spectral representations as
the Mandelstam representation and a strengthened form of crossing called
duality. The first one served to specify analyticity domains and the second
implemented the purely phenomenological idea of saturating the coexistence
of charges in the different crossed channels already by disregarding cuts
and only taking one-particle poles into account (``Nuclear Democracy'',
``Reggeization''). I believe that these added requirements which made heavy
use of analyticity contributed to the failure of the program. It did not
even achieve to reproduce those perturbative S-matrices which via LSZ
scattering theory were obtained from the Feynman perturbation theory of
time-ordered functions. What was left over from this program got swept aside
by the ascending gauge theory at the beginning of the 70$^{ies}.$

For the purpose of a good understanding of the content of this paper it is
very helpful to localize the cause of the failure of the S-matrix bootstrap
program. I think I will be in agreement with most of my colleagues who
followed these developments or later red about them that free-floating (and
often ill-defined) analyticity requirements are too fine instruments in
order to harmonize with physical intuition. Only analytic properties which
appear directly in the formulation of physical concepts are useful for the
construction of theories. This is best illustrated by two examples. The
x-space analyticity of correlation functions in QFT which was discovered by
Wightman \cite{St-Wi} is equivalent to the spectral-, covariance- and
locality- properties of the operator theory. On the other hand the
dispersion relations, even if restricted to the simplest case of forward
scattering, involve analyticity properties which arise from a quite
complicated interplay between the off-shell causality of retarded functions
with on-shell spectrum properties \cite{Todor}. Such non-constructive
analytic properties are still useful for experimentally verifying particular
consequences of causality and they also have their merits in the study of
possible nonperturbative high energy bounds on cross sections, but they have
no natural role in an actual construction.

Only a very few people took notice of the fact that the bootstrap program
finally worked in the more limited context of d=1=1 integrable models; it
was too far away from a ``theory of everything' which was on the minds of
the ambitious protagonists of the 4-dim. bootstrap which finally ended in
failure. The modest 2-dim. program led to a nice classification of families
of factorizing elastic S-matrices (thus showing that the idea of the
bootstrap being a TOE was incorrect ) and it also set the path for the
construction of associated QFT models via a formfactor program. A side
result of the S-matrix research in d=1+1 was the discovery of an on-shell
perturbation theory which, if specialized to on-shell tree graphs without
particle creation\footnote{%
The absense of particle creation is not an issue which is evident on the
level of tree graphs since it only happens happens on-shell. Properties
which are only valid on-shell are too subtle to be seen by inspection of
Feynman diagrams.} allowed to show the absence of creation for on-shell
one-loop approximation \cite{Berg}. Apparently the extension to multi-loops
was never elaborated in sufficient generality; some partial results can be
found in \cite{Weisz}. The very existence of these formulae shows that a
finite on-shell approach which avoids the characteristic off-shell short
distance problems of QFT is more than just a nice dream.

The present line of research directly takes off where the original program
failed. It removes the unfortunate TOE ideology\footnote{%
``Theories of everything'' seem to be also the favorite pass time of \ post
S-matrix physicist. The underlying idea that certain principles allow for
only one solution usually originate in connection with nonlinear structures
for which initially no solution is known.} from the S-matrix bootstrap and
incorporates the latter with the help of modular theory into the QFT
structure of wedge algebras. This return into QFT is based on the fact that
the S-matrix has in addition to the large time scattering interpretation
(well-known from the LSZ theory) another little known aspect namely it is
the relative modular invariant between the wedge algebra of incoming free
fields and that of the actual interacting Heisenberg operators. Whereas the
scattering aspect also applies to QM, the modular role is totally
characteristic for local quantum physics. Having established a direct
modular relation between the S-matrix and the wedge algebra, the old
S-matrix formalism becomes enriched with new physical concepts and
mathematical tools. In particular the thermal aspects of the wedge algebra
(the Hawking-Unruh temperature of matter behind a Rindler horizon) gets
inexorably tied up with the crossing symmetry of particle physics. A pivotal
role in the linkage of the S-matrix with the wedge algebras is played by
special wedge-localized operators which applied to the vacuum create
one-particle states without the usually associated cloud of
particle-antiparticles well known from the vacuum polarization phenomenon.
These PFGs (polarization-free generators) cannot exist for spacetime
localization regions whose causal completion is smaller than a wedge, unless
the theory has no interactions, in other words the wedge region is the
smallest causally complete region for which PFGs are compatible with the
presence of interactions.

The new framework shares with the old S-matrix bootstrap program (and with
string theory) the absence of any ultraviolet problems since it uses no
coordinatizations in terms of pointlike fields. Whether a theory exists or
not is not decided by the short distance singularities of some field
coordinates in terms of which the Lagrangian quantization happened to be
done, but rather depends on the nontriviality of the intersection structure
of wedge algebras. If intersections representing double cone algebras
contain more operators than just multiples of the identity, the theory is
nontrivial in the sense that possesses a nontrivial net for small
localization regions and not just for wedge regions. This avoidance of
particular pointlike fields and their short-distance problems was the main
dream and the raison d'etre of the S-matrix bootstrap. It is fully realized
in the new approach by the use of the field-coordinate independent algebraic
formulation of QFT (AQFT). The intention of the S-matrix protagonists to
abandon fields was reasonable, but unfortunately they thought that they also
should abandon the principle of locality.

The philosophy that the S-matrix has nothing to do with localization and
locality was also not quite right, as the relation to wedge-localized
algebras shows. Since from the net of wedge algebras one can get the
algebras of compact regions by intersections, all of local quantum physics
is determined by the S-matrix. An AQFT for a given S-matrix turns out to be
uniquely determined and it is believed that if S is admissable in the old
sense (unitary, crossing symmetric+associated analyticity), the net of wedge
algebras also exist and leads to the required nontrivial intersections. For
d=1+1 factorizing S-matrices the formfactor program goes a long way towards
proving this conjecture.

After having outlined our physical motivation and the position of the new
concepts with respect to older ideas, we now briefly mention our main
mathematical tool which will be used for problems of (quantum) localization:
(Tomita's) modular theory of von Neumann algebras\footnote{%
Special aspects (the thermal KMS characterization) of Tomita's mathematical
modular theory were discoverd by physicists in connection with quantum
statistics of Fermi/Bose systems formulated directly in the infinite volume
limit when the Gibbs formulation breaks down \cite{Ha}.}. These concepts,
which for the first time clarified the on/off-shell relation and in
particular the spacetime interpretation of on-shell crossing symmetry, were
not available at the time of the S-matrix bootstrap of the 60$^{ies}.$ In a
seminal paper \cite{Bi-Wi} the connection of wedge localized algebras with
modular theory was for the first time established. The present approach may
be considered as the inverse of the Bisognano-Wichmann Theorem. Instead of
extracting a deep mathematical property from the QFT of wedge algebras, we
are using this property together with the modular role of the S-matrix for
the construction of QFTs via wedge algebras. The main new mathematical tool
is briefly described in an appendix, a more detailed account can be found in 
\cite{Bo1}.

The ideas of AQFT used in this paper are not as much known as their
importance would suggest. Perhaps this is due to the fact that most particle
physicist consider QFT as a basically settled issue with only some nasty
technical problems remaining. We will demonstrate in this paper that such a
view is quite premature and unrealistic.

We have organized this paper as follows. The next section reviews and
illustrates the field-coordinate-free approach for theories without
interactions and for interacting d=1+1 factorizing model. In the latter case
the Zamolodchikov-Faddeev algebra emerges in a natural way without having
been put in \cite{Sch1}, and the hitherto formal Z-F operators for the first
time acquire a spacetime interpretation in connection with the new PFG
generators of wedge algebras. The presentation of these polarization-free
wedge generators is extended to systems which are not factorizing (i.e. to
theories with on-shell particle creation) in section 3.

After a brief introduction to the AQFT framework in section 4, the fifth
section treats the light ray/front restriction and algebraic holography in
terms of associated chiral conformal field theories. This connection is
again a deep result of modular theory, more specifically of modular
inclusions and intersections. There we also discuss the problem of inverting
such maps (the ``blow up'' property) in such a way that the original theory
becomes reconstructed from a finite number of copies of one abstract chiral
theory whose relative position in one Hilbert space has to be carefully
chosen. In colloquial terms this is like scanning a higher dimensional
massive theory by one chiral theory in different positions and we will refer
to it as ``chiral scanning''. Since a finite number of relatively positioned
chiral theories seems to be easier understandable than one higher
dimensional massive theory, the chiral scanning is in addition to the wedge
algebra method and the use of PFGs explained before a second potential
constructive idea based on modular theory. The mathematical technology used
in this section 5 is one of the most powerful which AQFT presently is able
to offer namely the theory of modular inclusions and intersections \cite
{Wies}\cite{Bo1}.

In the same section 5 we also take up the problem of associating entropy for
localized matter. In view of the fact that modular localization leads to
thermal aspects which show up in the appearance of a Hawking-Unruh
temperature it is natural to ask for a concept of ``localization entropy''
and in case it exists whether it gives a quantum version of the Bekenstein
area law for the area of the causal horizon of the localization region.

The last section finally tries to compare our approach with string theory.
This is on the one hand natural since both have similar historical roots,
but difficult from a conceptual viewpoint because string theory despite all
its mathematical formalism has never developed beyond a collection of
recipes towards formulating underlying principles. Whereas our approach has
strong ties with the older S-matrix bootstrap program (minus the wrong TOE
philosophy) which only used properties abstracted from QFT, string theory
via the dual model has added many ad hoc inventions which did not originate
through the intrinsic logic of S-matrix theory and QFT and are not asked for
by any known principle of particle physics. This is in my view the reason
why string theory despite all its semantic changes has remained a collection
of computational prescriptions without the guidance of a conceptual
framework. Since basic physical issues as locality and localizations,
algebras versus states etc. should be discussed on the level of physical
principles and not by looking at formalism and computational recipes, our
compression in the last section unavoidably remains somewhat vague and
superficial.

This presentation is in part a survey of published material \cite{Sch1}\cite
{S-W1}\cite{Essay} \cite{S-W2}\cite{S-W3}\cite{GLRV} as well as of new
results, in particular the presentation of localization-entropy in section
4. For the convenience of the reader we attached an two part appendix the
first part containing standard facts about modular theory and the second
proving that in interacting theories there are no polarization-free
generators for subwedge spacetime regions.

\section{Systems without Interactions and Factorizing Models}

In trying to bring readers with a good knowledge of standard QFT in contact
with some new (and old) concepts in algebraic QFT (AQFT) without sending him
back with a load of homework, I face a tricky problem. Let us for the
time-being put aside the intrinsic logic which would ask for a systematic
presentation of the general framework, and let us instead try to maneuver in
a more less ad hoc (occasionally even muddled) way.

In a pedestrian approach the problem of constructing nets of interaction
free systems from Wigner's one particle theory may serve as a nice
pedagogical exercise for the new ideas. The reader who is not familiar with
Wigner's representation theoretical method to describe particle spaces is
referred to \cite{Wei}\cite{Ha}. Since Wigner's representation theory
restricted to positive energy representations was the first totally
intrinsic relativistic quantum theory without any quantization parallelism
to classical particle theory, it is reasonable to expect in general that if
we find the right concepts, we should be able to avoid covariant pointlike
fields altogether in favor of a more intrinsic way to implement the
causality/locality principle. In that case the local fields should be
similar to coordinatizations of local observables in analogy to the use of
coordinates in modern differential geometry. This viewpoint is indeed
consistent and essential in the present context\cite{Sch1}\cite{BGL}.

Let us first understand how the free field algebras are directly abstracted
from the Wigner theory. By using a spatial variant of Tomita's theory for
the wedge situation i.e. by defining a kind of antilinear involutive
``pre-Tomita'' operator $\frak{s}$ on the Wigner representation space
(without a von Neumann algebra), one obtains a real closed subspace $%
H_{R}(W) $ of the Wigner space $H$ of complex multi-component momentum space
wave functions as a +1 eigenspace of the Tomita-like quantum mechanical
operator $\frak{s}$ in $H.$ Here $W$ denotes the x-t wedge $x>\left|
t\right| $ and $\frak{s}$ is defined to be the product of the $i\pi $%
-continued x-t Lorentz boost $\delta ^{\frac{1}{2}}$(obtained by the
functional calculus associated with the spectral theory of the boost
operator $\delta ^{i\tau }:=U(\Lambda _{x,t}(\pi \tau )),$ $\pi \tau =$%
rapidity) multiplied with the one-particle version of the (antiunitary) $j$%
-reflection\footnote{%
Apart from a rotation around the x-axis by an angle $\pi ,$ this is the
famous TCP operator restricted to the one-particle/antiparticle subspace.}
in the x-t plane 
\begin{eqnarray}
\frak{s} &=&j\delta ^{\frac{1}{2}} \\
s\psi &=&\psi ,\,\,\psi \in H_{R}(W)  \notag
\end{eqnarray}
For the definition of the antiunitary Tomita involution $j$ which represents
the x-t reflection in case of antiparticles $\neq $ particles) one needs to
extend the Wigner representation to the direct sum of particle/antiparticle
spaces; a process well-known in the Wigner theory if one wants to include
the disconnected Poincar\'{e} transformations. Since the x-t reflection
commutes with the x-t boost $\delta ^{it}$ and is antiunitary, it formally
inverts the unbounded $\delta $ i.e. $j\delta =\delta ^{-1}j$ which is
formally the analytically continuing boost at the imaginary value $t=-i.$ As
a result of this commutation relation the unbounded antilinear operator $%
\frak{s}$ is involutive on its domain of definition $\frak{s}^{2}\subset 1.$
These unusual properties, which are not met anywhere else in QM, encodes
geometric localization properties within abstract operator domains \cite
{Sch1} \cite{BGL}. They also preempt the relativistic locality properties of
QFT which Wigner looked for in vain \cite{New-Wi}. The opposite localization
i.e. $H_{R}(W^{opp})$ turns out to correspond to the symplectic (or real
orthogonal) complement of $H_{R}(W)$ in $H$ i.e. $\func{Im}(\psi
,H_{R}(W))=0\curvearrowright \psi \in H_{R}(W^{opp}).$ One furthermore finds
the following properties for the subspaces called ``standardness'' 
\begin{eqnarray}
&&H_{R}(W)+iH_{R}(W)\,\,is\,\,dense\,\,in\,\,H \\
&&H_{R}(W)\cap iH_{R}(W)=\left\{ 0\right\}  \notag
\end{eqnarray}
Having arrived at the wedge localization spaces, one may construct
localization spaces for smaller spacetime regions by forming intersections
over all wedges which contain this region 
\begin{equation}
H_{R}(\mathcal{O})=\bigcap_{W\supset \mathcal{O}}H_{R}(W)  \label{int}
\end{equation}
These spaces are again standard and have their own premodular objects $%
\delta ,j$ and $\frak{s,}$ but this time their action cannot be described in
terms of spacetime diffeomorphism. Note also that the modular formalism
characterizes the localization of subspaces, but is not able to distinguish
individual elements in that subspace. There is a good physical reason for
that, because as soon as one tries to do that, one is forced to leave the
unique Wigner $(m,s)$ representation framework and pick a particular
covariant representation by selecting one specific intertwiner among the 
\textit{infinite set} of $u$ and $v$ intertwiners which link the unique
Wigner (m,s) representation to the countably infinite many covariant
possibilities \cite{Sch1}. In this way one would then pass to the framework
of covariant fields explained and presented in the first volume of
Weinberg's book \cite{Wei}. The description of a concrete element in $%
H_{R}(W)$ or $H_{R}(\mathcal{O})$ then depends on the choice of covariant
formalism. A selection by e.g. invoking Euler equations and the existence of
a Lagrangian formalism may be convenient for doing particular perturbative
computations or as a mnemotechnical device for classifying polynomial
interaction densities\footnote{%
The causal approach permits the transformation of a polynomial interaction
from one coordinatization to any other whereas a formalism using classical
actions involving free field Lagrangians $\mathcal{L}_{0}$ is restricted to
the use of Euler-Lagrange field coordinatizations.}, but is not demanded as
an intrinsic attribute of physics.

The way to avoid nonunique covariant fields is to pass from Wigner subspaces
directly to von Neumann subalgebras of the algebra of all operators in Fock
space $B(\mathcal{H}_{Fock}$), i.e. the transition from real subspaces to
von Neumann subalgebras in Fock space is well known. With the help of the
Weyl (or CAR in case of Fermions) functor $Weyl(\cdot )$ one defines the
local von Neumann algebras \cite{Sch1}\cite{BGL} generated from the Weyl
operators 
\begin{equation}
\mathcal{A}(W):=alg\left\{ Weyl(f)|f\in H_{R}(W)\right\}  \label{Weyl}
\end{equation}
a process which is sometimes misleadingly called ``second quantization''.
These Weyl generators have the formal appearance 
\begin{eqnarray}
Weyl(f) &=&e^{ia(f)} \\
a(f) &=&\sum_{s_{3}=-s}^{s}\int (a^{\ast }(p,s_{3})f_{s_{3}}(p)+h.c.)\frac{%
d^{3}p}{2\omega }  \notag
\end{eqnarray}
i.e. unlike the covariant fields they are independent of the nonunique
intertwiners and depend solely on the unique Wigner data. An analogue
statement holds for the halfinteger spin case for which the CAR functor maps
the Wigner wave function into the fermionic generators of von Neumann
subalgebras. The statistics is already preempted by the premodular theory on
Wigner space \cite{Sch1}. The local net $\mathcal{A}(\mathcal{O})$ may be
obtained in two ways, either one first constructs the spaces $H_{R}(\mathcal{%
O})$ via (\ref{int}) and then applies the Weyl functor, or one first
constructs the net of wedge algebras (\ref{Weyl}) and then intersects the
algebras.

If we would have taken the conventional route via interwiners and local
fields as in \cite{Wei}, then we would have been forced to use Borchers
construction of equivalence classes\footnote{%
The class of covariant free fields belonging to the same (m,s) is a linear
subclass of the full equivalence class which comprises all Wick-polynomials.
In the analogy with coordinates in differential geometry this subclass
corresponds to linear coordinate transformations.} in order to see that the
different free fields associated with the $(m,s)$ representation with the
same momentum space creation and annihilation operators in Fock space are
just different generators of the same coherent families of local algebras
i.e. yield the same net. This would be analogous to working with particular
coordinates in differential geometry and then proving at the end that the
objects of interests are invariant and therefore independent of coordinates.

The implementation of interactions in the framework of nets requires a
radical re-thinking of the formalism, even if we are only interested in
perturbative aspects. The use of the above method for the Wigner
one-particle representation and the subsequent introduction of interactions
will inevitably force us to reintroduce field coordinates in order to define
what we mean by perturbative interactions. In order to avoid the standard
approach we therefore have to find a way to introduce interactions directly
into wedge-localized multiparticle spaces or wedge algebras.

In order to get a clue of how we can avoid the use of pointlike fields in
interacting situations, let us first ask this question in a more limited
context. It is well-known that there exists a special class of theories in
d=1+1 in which the S-matrix commutes with the incoming particle number

\begin{equation}
\left[ S_{sc},\mathbb{N}_{in}\right] =0
\end{equation}
and factorizes on multi-particle in-states \cite{STW}\cite{BKTW}\cite{Zam}.
For this reason these theories are often referred to as factorizing or
integrable (since this leads to an infinite number of conservation laws)
models. For those one finds that not only the old bootstrap program can be
carried through, but the application of the so-called formfactor-program
allows to compute even the fields in the sense of bilinear forms between in
and out states \cite{KW}\cite{Smir}. Let us ignore those
bootstrap-formfactor recipes and try find a modular access to these models
by implementing the idea of a relativistic particle pair interaction with a
most naive Ansatz (assuming for simplicity a situation of selfconjugate
particles) which formally generalizes the standard creation/annihilation
operators.\ Using rapidities instead of momenta we require 
\begin{eqnarray}
Z(\theta )Z(\theta ^{\prime }) &=&S(\theta -\theta ^{\prime })Z(\theta
^{\prime })Z(\theta )  \label{Z-F} \\
Z(\theta )Z^{\ast }(\theta ^{\prime }) &=&S^{-1}(\theta -\theta ^{\prime
})Z^{\ast }(\theta ^{\prime })Z(\theta )+\delta (\theta -\theta ^{\prime }) 
\notag
\end{eqnarray}
with the star-structure determining the remaining commutation relations and
the unitarity of $S$ with $S^{-1}(\theta )=\bar{S}(\theta )=S(-\theta )$
etc. Together with $Z(\theta )\Omega =0$ we can express all $Z$ correlation
functions in terms of $S$'s and the computation of correlation functions
proceeds as for free fields namely by commuting the annihilation operators $%
Z $ to the right vacuum e.g. 
\begin{eqnarray}
\left( \Omega ,Z(\theta _{4})Z(\theta _{3})Z^{\ast }(\theta _{2})Z^{\ast
}(\theta _{1})\Omega \right) &=&S(\theta _{2}-\theta _{3})\delta (\theta
_{3}-\theta _{1})\delta (\theta _{4}-\theta _{2})  \label{4} \\
&&+\delta (\theta _{3}-\theta _{2})\delta (\theta _{4}-\theta _{1})  \notag
\end{eqnarray}
Although we use the preemptive notation ``$Z"$ which refers to the
Zamolodchikov-Faddeev algebra\footnote{%
The missing delta-function contribution in Zamolodchikovs original proposal 
\cite{Za} was later added by Faddeev.}, there are for the time being no
requirements on the coefficients which go beyond those following from the
structure of a distributive $^{\ast }$-algebra, i.e. crossing symmetry is
not required but will be result from wedge localization.

One easily sees that instead of postulating commutation relations we could
also have started from the following formula which represents $Z^{\ast
}Z^{\ast }\Omega $ state vectors in terms of corresponding free field terms 
\begin{equation}
Z^{\ast }(\theta _{2})Z^{\ast }(\theta _{1})\Omega =\frac{1}{\sqrt{2}}(\chi
_{21}a^{\ast }(\theta _{2})a^{\ast }(\theta _{1})\Omega +\chi _{12}S(\theta
_{2}-\theta _{1})a^{\ast }(\theta _{1})a^{\ast }(\theta _{2})\Omega )
\label{rule}
\end{equation}
Here the symbol $\chi _{P(1)...P(n)}$ denotes the characteristic function of
the region $\theta _{P(1)}>...>\theta _{P(n)}.$ It is easy to see that the
inner product agrees with (\ref{4}); one only has to use the identity 
\begin{eqnarray}
&&\left\{ \chi _{12}S(\theta _{2}-\theta _{1})+\chi _{21}\bar{S}(\theta
_{3}-\theta _{4})\right\} \delta (\theta _{3}-\theta _{1})\delta (\theta
_{4}-\theta _{2})  \label{4-point} \\
&=&S(\theta _{2}-\theta _{1})\delta (\theta _{3}-\theta _{1})\delta (\theta
_{4}-\theta _{2})  \notag
\end{eqnarray}
In fact if we had started with a more general two-particle interaction
Ansatz by allowing the structure of the second equation in (\ref{Z-F}) to be
different say $S^{-1}\rightarrow T,$ the consistency with (\ref{rule}) would
immediately force us to return to $T=S^{-1}.$

The formula for the 4-point function suggests the possibility to replace the
algebraic Ansatz by the following formula for multi-$Z^{\ast }$ state
vectors 
\begin{eqnarray}
&&Z^{\ast }(\theta _{n})...Z^{\ast }(\theta _{1})\Omega  \label{state} \\
&=&\sum_{perm}\chi _{P(n)...P(1)}(\prod_{transp}S)a^{\ast }(\theta
_{P(n)})...a^{\ast }(\theta _{P(1)})\Omega  \notag
\end{eqnarray}
where the product of S-factors in the bracket contains one S for each
transposition which expresses the two-body nature of the interaction. The
associativity of the $Z^{\prime }s$ i.e. the Yang-Baxter relation for
matrix-valued $S^{\prime }s$ insures the consistency of the formula. We call 
$\theta _{P(1)}>...>\theta _{P(n)}$ the natural order of the multi-$Z^{\ast
} $ state vector. From the state characterization (\ref{state}) one can
derive the algebraic definition (\ref{Z-F}).

With these algebraic prerequisites out of the way, let us now return to
physics and investigate spacetime localization properties of the following
hermitian operators 
\begin{eqnarray}
&&F(\hat{f})=\int F(x)\hat{f}(x)d^{2}x,\,\,\,supp\hat{f}\,\in W  \label{path}
\\
&=&\int_{C}Z(\theta )\bar{f}(\theta ),\,\,\,\,Z(\theta -i\pi ):=Z^{\ast
}(\theta )  \notag \\
&&\hat{f}(x)=\frac{1}{\sqrt{2\pi }}\int (f(\theta )e^{-ipx}+c.c.)d\theta
,\,\,\bar{f}(i\pi -\theta )=f(\theta )  \notag
\end{eqnarray}
where C is a path consisting of the upper/lower rim of a $i\pi $-strip with
the real $\theta $-axis being the upper boundary. Whereas the on-shell value
of the Fourier transform $f(\theta )$ of $\hat{f}$ is analytic in this
strip, the last relation is a notation (since operators by themselves are
never analytic in spacetime labels!) which however inside expectation values
becomes coherent with meromorphic properties. If we take instead of $Z^{\#}$
free creation/annihilation operators, the corresponding formula $%
\int_{C}a(\theta )\bar{f}(\theta )$ represents wedge-localized smeared free
fields. Formally we may write in analogy to free fields 
\begin{eqnarray}
F(x) &=&\frac{1}{\sqrt{2\pi }}\int (e^{-ipx}Z(\theta )+h.c.)d\theta \\
p &=&m(cosh\theta ,sinh\theta )  \notag
\end{eqnarray}
but we should be aware that the argument x is not related to a pointlike
localization in the sense of causality since on-shell fields are local iff
they are bona fide free fields i.e. iff the $Z^{\prime }s$ reduce to the
standard creation/annihilation operators (see appendix).

In the following we will prove that the operators $F(\hat{f})$ with $Z^{\#}$
fulfilling (\ref{Z-F}) are localized in the wedge $x>\left| t\right| $ if
and only if the $^{\ast }$-algebra can be extended to a
Zamolodchikov-Faddeev algebra i.e. iff the coefficients S are crossing
symmetric including the crossing symmetric bootstrap pole structure.
Following our previously introduced terminology \cite{Sch1}, we will use the
name \textbf{p}olarization-\textbf{f}ree \textbf{g}enerators (PFG) for
localized operators in interacting QFT whose one time application to the
vacuum vector results in a one particle state vector. It is well-known \cite
{Mund} that PFGs with smaller than wedge localization regions (e.g. double
cones, spacelike cones) can only exist in theories without interactions i.e. 
$\psi _{f_{1}...f_{n}}^{in}=\psi _{f_{1}...f_{n}}^{out}.$ For the
convenience of the reader we present the argument in an appendix. PFGs
however always exist in regions in interacting theories if the localization
region is a wedge or bigger \cite{BBS}. The argument is based on modular
theory and will be recollected in the next section.

We want to show that the above $F$ can be converted indeed into bona fide
PFGs and a for a proof we have to check the KMS\ property for the
F-correlation functions with the modular generator being the infinitesimal
boost K. This property is a prerequisite for any wedge-localized algebra in
a Wightman QFT \cite{Bi-Wi}. The KMS property is well-known from statistical
mechanics and is the substitute for Gibbs formula which for many quantum
systems becomes meaningless in the thermodynamic limit. In the present
context its thermal aspects has been discussed in \cite{Sch1}. The desired
KMS-property for the wedge reads 
\begin{equation}
\left\langle F(\hat{f}_{n})...F(\hat{f}_{1})\right\rangle =\left\langle F(%
\hat{f}_{n-1})...F(\hat{f}_{2})F(\hat{f}_{n}^{2\pi i})\right\rangle
\end{equation}
where the superscript $2\pi i$ indicates the imaginary rapidity translation
from the lower to the upper rim of the KMS strip.

A rather straightforward calculation based on the previously explained rules
for the Zs yields the following result

\begin{theorem}
(\cite{Sch1}\cite{S-W2}) the KMS-thermal aspect of the wedge algebra
generated by the PFGs is equivalent to the crossing symmetry of the S-matrix 
\begin{equation*}
\mathcal{A}(W):=alg\left\{ F(\hat{f});supp\hat{f}\in W\right\}
\Leftrightarrow S(\theta )=S(i\pi -\theta )
\end{equation*}
Furthermore the possible crossing symmetric poles in the physical strip of S
will be converted into intermediate composite particle states in the GNS
Hilbert space associated with the state defined by the correlations on the $%
\mathcal{A}$($W$)-algebra. The latter commutes with its geometric opposite $%
\mathcal{A}$($W^{opp})$ in case of $\mathcal{A}$($W^{opp})=\mathcal{A}%
(W)^{\prime }=AdJ\mathcal{A}(W).$ A sufficient condition for this is the
existence of a parity transformation whose action on $\mathcal{A}(W)$ equals
the commutant $\mathcal{A}(W)^{\prime }.$
\end{theorem}

Since the Fs are unbounded operators with (particle number) N-bounds which
are the same as for free fields, the algebra generated by them is to be
understood in the sense that they are affiliated with that von Neumann
algebra which they generate.

We recall the proof for the 4-point function of $F^{\prime }s$ which may be
obtained as the scalar product of two-particle state vectors ($c.t.$ denotes
the F-contraction terms)

\begin{eqnarray}
&&F(\hat{f}_{2})F(\hat{f}_{1})\Omega =\int \int \bar{f}_{2}(\theta _{2}-i\pi
)\bar{f}_{1}(\theta _{1}-i\pi )Z^{\ast }(\theta _{1})Z^{\ast }(\theta
_{2})\Omega +c.t. \\
&=&\int \int \bar{f}_{2}(\theta _{2}-i\pi )\bar{f}_{1}(\theta _{1}-i\pi
)\{\chi _{12}a^{\ast }(\theta _{1})a^{\ast }(\theta _{2})\Omega +  \notag \\
&&+\chi _{21}S(\theta _{2}-\theta _{1})a^{\ast }(\theta _{2})a^{\ast
}(\theta _{1})\Omega \}+c\Omega
\end{eqnarray}
and the analogous formula for the bra-vector. The formula needs some
explanation. The symbol $\chi $ with the permutation subscript denotes as
before the characteristic function associated with the permuted rapidity
order. The order for the free creation operators $a^{\ast }$ is governed by
particle statistics. For each transposition starting from the natural order (%
\ref{state}), one obtains an $S$ factor\footnote{%
The notation has used the statistics in order to bring the product of
incoming fields $a_{in}$ into the natural order say 1...n. The ordering of
the $Z^{\prime }s$ encodes the $\theta $-ordering and not the particle
statistics. It is connected with the different boundary values of state
vectors and expectation values in $\theta $- space in approaching the
physical boundary from the analytic region. This is analogous to the
association of the n! n-point x-space correlation functions with different
boundary values of one analytic ``master function'' in the Wightman theory.}%
. The Yang-Baxter relation assures that the various ways of doing this are
consistent. For the inner product the $S$-dependent terms are. Finally the
terms proportional to the vacuum are contraction terms corresponding to the $%
\delta $-function in (\ref{Z-F}). For the $S$-dependent terms in the inner
product we obtain 
\begin{eqnarray}
&&\int \int f_{4}(\theta _{2})f_{3}(\theta _{1})\{\chi _{21}S(\theta
_{2}-\theta _{1})+\chi _{12}\bar{S}(\theta _{1}-\theta _{2})\}\bar{f}%
_{2}(\theta _{2}-i\pi )\bar{f}_{1}(\theta _{1}-i\pi )d\theta _{1}d\theta _{2}
\notag \\
&=&\int \int f_{4}(\theta _{2})f_{4}(\theta _{1})S(\theta _{2}-\theta _{1})%
\bar{f}_{2}(\theta _{2}-i\pi )\bar{f}_{1}(\theta _{1}-i\pi )d\theta
_{1}d\theta _{2}  \label{KMS1}
\end{eqnarray}

The analogous computation for KMS crossed term in (\ref{KMS}) gives 
\begin{equation}
\int \int f_{2^{\prime }}(\theta _{1})f_{2}(\theta _{2})S(\theta _{1}-\theta
_{2})f_{1}(\theta _{1}-i\pi )f_{1^{\prime }}(\theta _{2}-i\pi +2\pi
i)d\theta _{1}d\theta _{2}  \label{KMS2}
\end{equation}
This formula makes only sense if the $F(f)$ operators are restricted in such
a way that the $2\pi i$ translation on them is well-defined, i.e. for wave
functions $f$ which are analytic in a strip of size $2\pi i.$ It is
well-known that the KMS condition does not hold on all operators of the
algebra but rather on a dense set of suitably defined analytic elements \cite
{Br-Ro}. The S-independent terms which we have not written down are
identical to terms in the 4-point function of free fields They separately
satisfy the KMS property. What remains is to show the identity of (\ref{KMS1}%
) and (\ref{KMS2}). This is done by $\theta _{2}\rightarrow \theta _{2}-i\pi 
$ contour-shift in (\ref{KMS2}) without picking up terms from infinity.
Using the denseness of the wave functions one finally obtains 
\begin{equation}
S(\theta _{2}-\theta _{1})=S(\theta _{1}-\theta _{2}+i\pi )
\end{equation}
which is the famous crossing symmetry or the $z\longleftrightarrow -z$
reflection symmetry around the point $z_{0}=\frac{1}{2}i\pi .$ For non
self-conjugate situation the crossed particles are antiparticles and the S
on the right hand side has to be modified accordingly.

\ In physical terms we may say that the wedge structure of factorizing
models is that of a kind of relativistic quantum mechanic. This continues to
be true if the crossing symmetric S-matrix has poles in the physical strip.
In that case the above contour shift would violate the KMS property unless
one modifies the multi-$Z^{\ast }$ state vector formula (\ref{state}) by the
inclusion of bound states. For the case n=2 (\ref{rule}) this means 
\begin{eqnarray}
&&Z^{\ast }(\theta _{2})Z^{\ast }(\theta _{1})\Omega =\left( Z^{\ast
}(\theta _{2})Z^{\ast }(\theta _{1})\Omega \right) ^{scat}+ \\
&&+\left| \theta ,b\right\rangle \left\langle \theta ,b\left| Z^{\ast
}(\theta -i\theta _{b})Z^{\ast }(\theta +i\theta _{b})\right| \Omega
\right\rangle   \notag
\end{eqnarray}
The bracket with the superscript scat denotes the previous contribution (\ref
{rule}), whereas the second line denotes the bound state contribution. The
validity of the KMS property demands the presence of this term and
determines the coefficient; here $\theta _{b}$ is the imaginary rapidity
related to the bound state mass. For a detailed treatment which includes the
bound state problem, we refer to a forthcoming paper. We emphasize again
that it is the representation of the $F$-correlations in terms of the
S-matrix and the KMS property of these correlation functions, which via the
GNS construction converts the poles in the (possibly matrix-valued) function
S into the extension of the Fock space of the $a^{\prime }s$ by additional
free field operators. In this way the poles in numerical functions are
converted into the enlargement of Fock space in such a way that a few $%
Z^{\prime }s$ can describe many more particles. One may call the Z to be
``fundamental'' and the introduce new $Z_{b}$ and $F_{b}^{\prime }s;$ the
latter will however be operators which are already associated with the
original F-algebra. What needs an extension is the wedge algebra of \textit{%
incoming} fields. It is very important to note that this apparent quantum
mechanical picture is converted into LQP \textit{with vacuum polarization as
soon as we e.g. go to double cone localization}; this will be shown in the
sequel. The extension of the above proof beyond 4-point functions is left to
the reader.

\textit{With this theorem relating wedge localization via the thermal KMS
property to crossing symmetry, we have achieved the main goal of this
section: to show that the Zamolodchikov-Faddeev algebra which consists of \ (%
\ref{Z-F}) together with the crossing symmetry of its structure function has
a deep spacetime interpretation and an associated thermal KMS aspect. In
fact the simplest PFGs which fulfill conservation of real particle number
and have only elastic scattering (possible in d=1+1) are precisely the Z-F
algebra operators! In a moment we will see that these models have the full
interacting vacuum structure (virtual particle nonconservation) with respect
to operators from smaller localization regions (e.g. double cones), i.e. we
are dealing with a genuine interacting field theory (and not some
relativistic quantum mechanics). }

The KMS computation can be immediately extended to ``formfactors'' i.e.
mixed correlation functions containing in addition to Fs one generic
operator $A\in \mathcal{A}(W)$ so that the previous calculation results from
the specialization $A=1.$ This is so because the connected parts of the
mixed correlation function is related to the various $\left( n,m\right) $
formfactors (\ref{1}) obtained by the different ways of distributing n+m
particles in and out states. These formfactors are described by different
boundary values of one analytic master function which is in turn related to
the various forward/backward on shell values which appear in one mixed A-F
correlation function. We may start from the correlation function with one $A$
to the left and say n Fs to the right and write the KMS condition as 
\begin{equation}
\left\langle AF(\hat{f}_{n})...F(\hat{f}_{2})F(\hat{f}_{1})\right\rangle
=\left\langle F(\hat{f}_{1}^{2\pi i})AF(\hat{f}_{n})...F(\hat{f}%
_{2})\right\rangle  \label{A-KMS}
\end{equation}
The n-fold application of the Fs to the vacuum on the left hand side creates
besides an n-particle term involving n operators $Z^{\ast }$ to the vacuum
(or KMS reference state vector) $\Omega $ also contributions from a lower
number of $Z^{\ast \prime }s$ together with $Z$-$Z^{\ast }$ contractions. As
with free fields, the n-particle contribution can be isolated by
Wick-ordering\footnote{%
Note that as a result of the commutation relation (\ref{Z-F}), the change of
order within the Wick-ordered products will produce rapidity dependent
factors} 
\begin{equation}
\left\langle A:F(\hat{f}_{n})...F(\hat{f}_{2})F(\hat{f}_{1}):\right\rangle
=\left\langle F(\hat{f}_{1}^{2\pi i})A:F(\hat{f}_{n})...F(\hat{f}%
_{2}):\right\rangle  \label{KMS}
\end{equation}
Rewritten in terms of \ $A$-formfactors the n-particle scattering
contribution (using the denseness of the $f$($\theta ))$ reads as 
\begin{eqnarray}
&&\left\langle \Omega ,AZ^{\ast }(\theta _{n})...Z^{\ast }(\theta
_{2})Z^{\ast }(\theta _{1}-2\pi i)\Omega \right\rangle \\
&=&\left\langle \Omega ,Z(\theta _{1}+i\pi )AZ^{\ast }(\theta
_{n})...Z^{\ast }(\theta _{2})\Omega \right\rangle  \notag \\
&=&\left\langle Z^{\ast }(\theta _{1}-i\pi )\Omega ,AZ^{\ast }(\theta
_{n})...Z^{\ast }(\theta _{2})Z^{\ast }(\theta )\Omega \right\rangle  \notag
\end{eqnarray}
Here the notation suffers from the usual sloppiness of physicists notation:
the analytic continuation by $2\pi i$ refers to the correlation function and
not to the operators. For the natural order of rapidities $\theta
_{n}>..>\theta _{1}$ this yields the following crossing relation 
\begin{eqnarray}
&&\left\langle \Omega ,Aa_{in}^{\ast }(\theta _{n})...a_{in}^{\ast }(\theta
_{2})a_{in}^{\ast }(\theta _{1}-\pi i)\Omega \right\rangle \\
&=&\left\langle a_{out}^{\ast }(\theta _{1})\Omega ,Aa_{in}^{\ast }(\theta
_{n})...a_{in}^{\ast }(\theta _{2})\Omega \right\rangle  \notag
\end{eqnarray}
The out scattering notation on the bra-vectors becomes only relevant upon
iteration of the KMS condition since the bra $Z^{\prime }s$ have the
opposite natural order. The above KMS relation (\ref{KMS}) contains
additional information about bound states and scattering states with a lower
number of particles. The generalization to the case of antiparticles$\neq $%
particles is straightforward. More generally we see that the connected part
of the mixed matrix elements 
\begin{equation}
\left\langle a_{out}^{\ast }(\theta _{k})...a_{out}^{\ast }(\theta
_{1})\Omega ,Aa_{in}^{\ast }(\theta _{n})...a_{in}^{\ast }(\theta
_{k-1})\Omega \right\rangle  \label{formfactors}
\end{equation}
is related to $\left\langle \Omega ,AZ^{\ast }(\theta _{n})...Z^{\ast
}(\theta _{2})Z^{\ast }(\theta _{1})\Omega \right\rangle $ by analytic
continuation which a posteriori justifies the use of the name formfactors in
connection with the mixed A-F correlation functions.

The upshot of this is that such an $A$ must be of the form 
\begin{equation}
A=\sum \frac{1}{n!}\int_{C}...\int_{C}a_{n}(\theta _{1},...\theta
_{n}):Z(\theta _{1})...Z(\theta _{n}):  \label{series}
\end{equation}
where the $a_{n}$ have a simple relation to the various formfactors of $A$
(including bound states) whose different in-out distributions of momenta
correspond to the different contributions to the integral from the
upper/lower rim of the strip bounded by C, which are related by crossing.
The transcription of the $a_{n}$ coefficient functions into physical
formfactors (\ref{formfactors}) complicates the notation, since in the
presence of bound states there is a larger number of Fock space particle
creation operators than PFG wedge generators $F.$ It is comforting to know
that the wedge generators, despite their lack of vacuum polarization clouds,
nevertheless contain the full (bound state) particle content. The wedge
algebra structure for factorizing models is like a relativistic QM, but as
soon as one sharpens the localization beyond wedge localization, the field
theoretic vacuum structure will destroy this simple picture and replace it
with the appearance of the characteristic virtual particle structure which
separates local quantum physics from quantum mechanics.

In order to see by what mechanism the quantum mechanical picture is lost in
the next step of localization, let us consider the construction of the
double cone algebras as a relative commutants of shifted wedge (shifted by $%
a $ inside the standard wedge) 
\begin{eqnarray}
\mathcal{A}(C_{a}) &:&=\mathcal{A}(W_{a})^{\prime }\cap \mathcal{A}(W)
\label{loc} \\
C_{a} &=&W_{a}^{opp}\cap W  \notag
\end{eqnarray}
For $A\in \mathcal{A}(C_{a})\subset \mathcal{A}(W)$ and $F_{a}(\hat{f}%
_{i})\in \mathcal{A}(W_{a})\subset \mathcal{A}(W)$ the KMS condition for the
W-localization reads as before, except that whenever a $F_{a}(\hat{f}_{i})$
is crossed to the left side of $A,$ we may commute it back to the right side
since $\left[ \mathcal{A}(C_{a}),F_{a}(\hat{f}_{i})\right] =0.$ The
resulting relations are e.g. 
\begin{eqnarray}
&&\left\langle AF_{a}(\hat{f}_{1}):F_{a}(\hat{f}_{n})...F_{a}(\hat{f}%
_{2}):\right\rangle  \label{rec} \\
&=&\left\langle A:F_{a}(\hat{f}_{n})...F_{a}(\hat{f}_{2})F_{a}(\hat{f}%
_{1}^{2\pi i}):\right\rangle  \notag
\end{eqnarray}
Note that the $F_{a}(\hat{f}_{1})$ in the first line is outside the
Wick-ordering. Since it does neither act on the bra nor the ket vacuum, it
contains both frequency parts. The creation part can be combined with the
other Fs under one common Wick-ordering whereas the annihilation part via
contraction with one of the Wick-ordered Fs will give an expectation value
of one $A$ with $(n-2$) $F$s. Using the density of the Fs and going to
rapidity space we obtain (\cite{S-W1}) the so-called kinematical pole
relation 
\begin{equation}
Res_{\theta _{12}=i\pi }\left\langle AZ^{\ast }(\theta _{n})...Z^{\ast
}(\theta _{2})Z^{\ast }(\theta _{1})\right\rangle =2i\mathbf{C}%
_{12}\left\langle AZ^{\ast }(\theta _{n})...Z^{\ast }(\theta
_{3})\right\rangle (1-S_{1n}...S_{13})  \label{pole}
\end{equation}
Here the product of two-particle S-matrices results from commuting the $%
Z(\theta _{1})$ to the right so that it stands to the left of $Z^{\ast
}(\theta _{2}),$ whereas the charge congugation matrix $\mathbf{C}$ only
appears if we relax our assumption of self-congugacy.

I believe that the general issue of the shape of polarization clouds in
terms of their asymptoptotic (say incoming) particle content is intimately
related to the structure of the yet unknown modular automorphisms which
exist for each spacetime region.

This relation appears for the first time in Smirnov's axiomatic approach 
\cite{Smir} as one of his recipes; more recently it was derived as a
consequence of the LSZ formalism adapted to the factorizing model situation 
\cite{BFKZ}. In the present approach it has an apparently very different
origin: it is together with the Z-F algebra structure a consequence of the
wedge localization of the generators $F(\hat{f})$ and the sharpened double
cone locality (\ref{loc}) of $A$. The existence problem for the QFT
associated with an admissable S-matrix (unitary, crossing symmetric, correct
physical residua at one-particle poles) of a factorizing theory is the
nontriviality of the relative commutant algebra i.e. $\mathcal{A}(C_{a})\neq 
\mathbf{C}\cdot 1.$ Intuitively the operators in double cone algebras are
expected to behave similar to pointlike fields applied to the vacuum; namely
one expects the full interacting polarization cloud structure. For the case
at hand this is in fact a consequence of the above kinematical pole formula
since this leads to a recursion which for nontrivial two-particle S-matrices
is inconsistent with a finite number of terms in (\ref{series}). Only if the
bracket containing the S-products vanishes, the operator $A$ is a composite
of a free field.

The determination of a relative commutant or an intersection of wedge
algebras even in the context of factorizing models is not an easy matter. We
expect that the use of the following ``holographic'' structure significantly
simplifies this problem. We first perform a \textit{lightlike translation}
of the wedge into itself by letting it slide along the upper light ray by
the amount given by the lightlike vector $a_{+}.$ We obtain an inclusion of
algebras and an associated relative commutant 
\begin{eqnarray}
&&\mathcal{A}(W_{a_{\pm }})\subset \mathcal{A}(W) \\
&&\mathcal{A}(W_{a_{\pm }})^{\prime }\cap \mathcal{A}(W)  \notag
\end{eqnarray}
The intuitive picture is that the relative commutant lives on the $a_{\pm 
\text{ }}$interval of the upper/lower light ray, since this is the only
region inside W which is spacelike to the interior of the respective shifted
wedges. This relative commutant subalgebra is a light ray part of the above
double cone algebra, and it has an easier mathematical structure. One only
has to take a generic operator in the wedge algebra which formally can be
written as a power series (\ref{series}) in the generators and \cite{Sch1} 
\cite{S-W2} find those operators which commute with the shifted Fs 
\begin{equation}
\left[ A,U(e_{+})F(f)U^{\ast }(e_{+})\right] =0
\end{equation}
Since the shifted Fs are linear expressions in the Zs, the $n^{th}$ order
polynomial contribution to the commutator comes from only two adjacent terms
in $A$ namely from $a_{n+1}$ and $a_{n-1}$ which correspond to the
annihilation/creation term in F. The size of the shift gives rise to a
Paley-Wiener behavior in imaginary direction, whereas the relation between $%
a_{n+1}$ and $a_{n-1}$ is identical to (\ref{pole}), so we do not learn
anything new beyond what was already observed with the KMS technique (\ref
{rec}). However as will be explained in section 5, the net obtained from the
algebra 
\begin{equation}
\mathcal{A}_{\pm }:=\cup _{a_{\pm }}A(C_{a\pm })
\end{equation}
is a chiral conformal net on the respective subspace $H_{\pm }=$ $\overline{%
\mathcal{A}_{\pm }\Omega }.$ If our initial algebra were d=1+1 conformal,
the total space would factorize $H=H_{+}\bar{\otimes}H_{-}=$ $\overline{%
\left( \mathcal{A}_{+}\bar{\otimes}\mathcal{A}_{-}\right) \Omega },$ and we
would recover the well-known fact that two-dimensional local theories
factorize into the two light ray theories. If the theory is massive, we
expect $H=\overline{\mathcal{A}_{+}\Omega }$ i.e. the Hilbert space obtained
from one horizon already contains all state vectors$.$ This would correspond
to the difference in classical propagation of characteristic massless versus
massive data in d=1+1. There it is known that although for the massless case
one needs the characteristic data on the two light rays, the massive case
requires only one light ray. In fact there exists a rigorous proof that this
classical behavior carries over to free quantum fields: with the exception
of m=0 massless theories, in all other cases (including light-front data for
higher dimensional m=0 situations) the vacuum is cyclic with respect to one
light front $H=\overline{\mathcal{A}_{\pm }\Omega }$ \cite{GLRV}. The proof
is representation theoretic and holds for all cases except the d=1+1
massless case. Hence in the case of interaction free algebras the
holographic light front reduction which, has d-1 dimensions, always fulfills
for d\TEXTsymbol{>}2 the Reeh-Schlieder property, where for d=1+1 only
massive theories obey holographic cyclicity. In order to recover the wedge
algebra with its net structure from the holographic restriction, one needs
the opposite light ray translation with $U(a_{-})$ i.e. $\mathcal{A}(W)=\cup
_{a_{-}<0}AdU(a_{-})\mathcal{A}_{+}$. For the nontriviality of the net
associated with $\mathcal{A}(W)$ it is sufficient to show that the
associated chiral conformal theory is nontrivial. In order to achieve this,
one has to convert the bilinear forms (\ref{series}) in the Z-basis which
fulfil the recursion relation into genuine operators on the one-dimensional
light ray. This is outside the scope of this paper.

Hence the modular approach leads to a dichotomy of \textit{real particle
creation} (absent in factorizing models) in the PFGs and in the aspect of
wedge localization, versus the full QFT \textit{virtual particle structure}
of the vacuum\footnote{%
The deeper understanding of the virtual vacuum structure (or the particle
content of say state vectors obtained by application of a double cone
localized operator to the vacuum) is presumably hidden in the modular groups
of double cone algebras.} for the more sharply localized operators. In some
sense the wedge is the best compromise between the particle/field point of
view. \textit{In this and only in this sense the particle-field dualism (as
a generalization of the particle-wave dualism of QM) applies to QFT}. Since
it is left invariant by an appropriate L-boost, the algebra contains enough
operators in order to resolve at least the vacuum and one-particle states
(which cannot be resolved from the remaining states in any algebra with a
smaller localization).

In the next section we will argue that these properties are not freaks of
factorizing models, whereas in a later section we will reveal the less
pedestrian aspects of light cone subalgebras and holography. As we have
argued on the basis of the previous pedestrian approach, the holography
aspect will be important in the modular construction of QFT's, because it
delegates certain properties of a rather complicated theory to those of (in
general several) simpler theories.

It is worthwhile to highlight two aspects which already are visible from
this pedestrian considerations. One is the notion of ``quantum
localization'' in terms of algebraic intersections as compared to the more
classical localization in terms of test function smearing of pointlike
fields. As mentioned already, the wedge localization of the PFGs cannot be
improved by choosing smaller supports of test functions inside the wedge;
the only possibility is to intersect algebras. In that case the old
generators become useless e.g. in the description of the double cone
algebras; the latter has new generators. Related to this is that the short
distance behavior looses its dominating (and somewhat threatening) role.

If one does not use field-coordinatizations, it is not even clear what one
means by ``the (good or bad) short distance behavior of a theory''. Short
distance behavior of what object? There is no short distance problem of
PFGs, since they have some ``natural cutoff'' (to the extend that the use of
such words which are filled with preassigned old meaning are reasonable in
the new context). \textit{Intersection of algebras does not give rise to
short distance problems} in the standard sense of this word. An explicit
construction of pointlike field coordinates from algebraic nets is presently
only available for chiral conformal theories \cite{Fr-Jo}. It produces
fields of arbitrary high operator dimension, and as a result of its group
theoretical techniques it also does not suffer from short distance problems.
The absence of short distance problems in the modular localization approach
seems to be of an entirely different nature than statements about the
absence of ultraviolet problems in string theory.

The results in this section should be viewed as an extension of the Wigner
theory into the realm of interactions for a special class of models.

\section{PFGs in Presence of Real Particle Creation}

For models with real particle creation it is not immediately clear how to
construct PFGs. In order to get some clue we first look at d=1+1 theories
which do not have any transversal extension to wedges. Furthermore we assume
that there is only one kind of particle which corresponds to the previous
assumption concerning the absence of poles in the two-particle S-matrix for
factorizing models. Modular theory always assures the existence of PFGs \cite
{BBS}. In fact for every modular localized state vector 
\begin{eqnarray}
S\psi &=&\psi  \label{modloc} \\
S &=&J\Delta ^{\frac{1}{2}}  \notag
\end{eqnarray}
There exists a (generally unbounded) operator $G$ associated with the von
Neumann $\mathcal{A}$ algebra in standard position such that 
\begin{eqnarray}
G\Omega &=&\psi \\
G^{\ast }\Omega &=&S\psi
\end{eqnarray}
For the case at hand $\mathcal{A}=\mathcal{A}(W),\Omega =\left|
0\right\rangle =$vacuum, there exists a dense subspace of explicitly
constructible (see below) wedge-localized state vectors $\mathcal{H}%
_{R}^{Fock}+i\mathcal{H}_{R}^{Fock}$ which possess affiliates Gs. Since $%
H_{R}(W)\subset \mathcal{H}_{R}^{Fock}$ there exists an algebra affiliated
operator $F(f)$ for each vector $f$ in $H_{R}(W)+iH_{R}(W)$ with 
\begin{eqnarray}
F(f)\Omega &=&\left| f\right\rangle =1-particle\,\,state \\
F(f)^{\ast }\Omega &=&S\left| f\right\rangle  \notag
\end{eqnarray}
Although the existence of PFGs outside of factorizing models poses no
problems, the presence of particle creation prevents them to have amenable
algebraic properties. The interpretation in the form of 
\begin{eqnarray}
F(f) &=&\int F(x)\hat{f}(x)d^{2}x \\
F(x) &=&\int \left( Z(\theta )e^{-ipx}+\bar{Z}^{\ast }(\theta
)e^{ipx}\right) d\theta
\end{eqnarray}
where $F(x)$ is a tempered operator-valued distribution on a dense
translation invariant domain which holds in the factorizing case is not
compatible with particle creation \cite{BBS} because it leads to relative
commutation relations of F with the incoming/outgoing free field 
\begin{eqnarray}
\left[ Z^{\#}(\theta ),a_{in}^{\#}(\theta ^{\prime })\right] &=&0,\,\,\theta
<\theta ^{\prime } \\
\left[ Z^{\#}(\theta ),a_{out}^{\#}(\theta ^{\prime })\right]
&=&0,\,\,\theta >\theta ^{\prime }  \notag
\end{eqnarray}
and similarly for the antiparticle operators $\bar{Z}^{\#}.$ Therefore we
will first try to see how far we can get with localized states.

Again we specialize to the self-conjugate case $\bar{Z}=Z$ and the absence
of bound states$.\,$From the previous discussion we take the idea that we
should look for a relation between the ordering of rapidities and the action
of the scattering operator. We fix the state vector $\psi (\theta
_{n},...,\theta _{1})$ for the natural $\theta $-order to be an incoming
n-particle state as we did for the previous particle conserving situation.
The totally mirrored order should then be a vector obtained by applying the
full S-matrix to the incoming n-particle vector.

But what should we do for the remaining permutations? We should end up with
a prescription which for factorizing systems agrees with the previous
formalism. For two $fs$ there is no problem; the formula looks as before (%
\ref{rule}), except that the application of the S-operator to the
two-particle in-vector has components to all n-particle multiparticle
vectors for $n\geq 2,$ i.e. the rapidities are labels which are not related
to the incoming particle content of the state vector

\begin{eqnarray}
&&\psi _{2}(f_{2},f_{1})\sim \iint_{C}(\chi _{21}a(\theta _{2})a(\theta
_{1})+\chi _{21}Sa(\theta _{1})a(\theta _{2}))\left| 0\right\rangle \bar{f}%
_{2}(\theta _{2})\bar{f}_{1}(\theta _{1})d\theta _{2}d\theta _{1}  \notag \\
&&\left\langle 0|a(\theta _{n})...a(\theta _{3})Sa^{\ast }(\theta
_{1})a^{\ast }(\theta _{2})|0\right\rangle \neq 0,\,\,n\geq 4
\end{eqnarray}
The check of the localization equation $J\Delta ^{\frac{1}{2}}\psi _{2}=\psi
_{2}$ with $J=J_{0}S$ (again we omitted the subscript $scat$ from the
S-matrix and we used the previous notation where the integration path $C$
includes creation as well as annihilation contributions there will be also a
contraction term. The inner product between two $\psi _{2}$ comes out to be

\begin{eqnarray}
&&\int \bar{f}_{1}^{^{\prime }}(\theta _{1}^{\prime })\bar{f}_{2}^{^{\prime
}}(\theta _{2}^{\prime })\left\langle 0|a(\theta _{1}^{\prime })a(\theta
_{2}^{\prime })a^{\ast }(\theta _{2})a^{\ast }(\theta _{1})|0\right\rangle
f_{2}(\theta _{2})f_{1}(\theta _{1})+c.t.  \label{2} \\
&&+\int \bar{f}_{1}^{^{\prime }}(\theta _{1}^{\prime })\bar{f}_{2}^{^{\prime
}}(\theta _{2}^{\prime })\left\langle 0\right| a(\theta _{2}^{\prime
})a(\theta _{1}^{\prime })(S\chi _{21}+S^{\ast }\chi _{12})a^{\ast }(\theta
_{2})a^{\ast }(\theta _{1}\left| 0\right\rangle f_{2}(\theta
_{2})f_{1}(\theta _{1})  \notag
\end{eqnarray}
where c.t. denote the contraction term contributions coming from an
annihilation part in $C$ and we have disregarded problems of overall
normalizations and the integration is done over all $\theta $ and $\theta
^{\prime }$. Using the unitarity property of S and the boundary property $%
\bar{f}(i\pi -\theta )=f(\theta ),$ the last term can be written without the
ordering $\chi s$ as 
\begin{equation}
\bar{f}_{1}^{^{\prime }}(\theta _{1}^{\prime })\bar{f}_{2}^{^{\prime
}}(\theta _{2}^{\prime })\left\langle 0\right| a(\theta _{2}^{\prime
})a(\theta _{1}^{\prime })Sa^{\ast }(\theta _{2})a^{\ast }(\theta _{1}\left|
0\right\rangle f_{2}(\theta _{2})f_{1}(\theta _{1})
\end{equation}
and has the same form as for the previous factorizing case if one replaces (%
\ref{KMS2}). In order to establish the KMS property we have to write this
inner product as 
\begin{equation}
(\psi _{2}^{\prime },\psi _{2})=(\psi _{1}^{\prime },\psi _{3})  \label{con}
\end{equation}
where $\psi _{1}$ is a one-particle vector. So we have to figure out how
permutations beyond the natural order and its mirror image are represented
on tensor product factors of incoming state vectors. Some thinking reveals
that subsequent applications of S-matrices on tensor factors of the
n-particle tensor product vectors only makes sense for nonoverlapping
situations. The action of the S-matrix on one tensor factor is associated
with the mirror perturbation of that tensor factor $12...k\rightarrow k...21$
since intuitively speaking one only obtains the full k-particle scattering
if the incoming velocities (or rapidities) are such that all particles meet
kinematically which only happens if the order of incoming velocities is the
mirrored natural order. Mathematically we should write each permutation as
the nonoverlapping product of ``mirror permutations'' The smallest mirror
permutations are transpositions of adjacent factors. An example for an
overlapping product is the product of two such transpositions which have one
element in common e.g. $123\rightarrow 132\rightarrow 312;$ there is no
meaning in terms of a subsequent tensor S-matrix action. However the
composition $123\rightarrow 213\rightarrow 312$ has a meaningful S-matrix
counterpart: namely $S\cdot S_{12}a^{\ast }(\theta _{1})a^{\ast }(\theta
_{2})a^{\ast }(\theta _{3})\Omega $ where $S_{12}$ leaves the third tensor
factor unchanged i.e. is the Fock space vector $\left( Sa^{\ast }(\theta
_{1})a^{\ast }(\theta _{2})\Omega \right) \otimes a^{\ast }(\theta
_{3})\Omega $ on which the subsequent action of $S$ (which corresponds to
the mirror permutation of all 3 objects) is well defined. In general if one
mirror permutation is completely inside a larger one the scattering
correspondence which is consistent with the tensor product structure of Fock
space. On the other hand for overlapping products of mirror permutations the
association to scattering data becomes meaningless, where overlapping means
that part of each mirror permutation is outside of the other. Fortunately,
as it is easy to see, there is precisely one representation in terms of
nonoverlapping mirror permutations. This leads to a unique representation of
multi $f$ labeled state vectors in terms of scattering data. On the other
hand if we were to write each mirror permutation as a product of
(necessarily overlapping) transpositions, we loose the uniqueness and we
then need the Yang-Baxter structure in order to maintain consistency; in
this case we return to the modular setting of factorizing models in the
previous section.

Let us elaborate this in a pedestrian fashion by writing explicit formulas
for n=3. The state vector is a sum of 3!=6 terms

\begin{multline}
\psi _{3}(f_{3},f_{2},f_{1})\simeq \int \int \int_{C}\{\chi _{321}a(\theta
_{3})a(\theta _{2})a(\theta _{1})+\chi _{312}S_{21}a^{\ast }(\theta
_{3})a^{\ast }(\theta _{2})a^{\ast }(\theta _{1}) \\
+\chi _{231}S_{32}a(\theta _{3})a(\theta _{2})a(\theta _{1})+\chi
_{123}S_{321}a(\theta _{3})a(\theta _{2})a(\theta _{1})  \notag \\
+\chi _{132}S_{321}\cdot S_{23}^{\ast }a^{\ast }(\theta _{3})a(\theta
_{2})a(\theta _{1})+\chi _{213}S_{321}\cdot S_{12}^{\ast }a(\theta
_{3})a(\theta _{2})a(\theta _{1})\}\left| 0\right\rangle \bar{f}_{3}(\theta
_{3})\bar{f}_{2}(\theta _{2})\bar{f}_{1}(\theta _{1})  \notag
\end{multline}

Here $\chi $ denotes again the characteristic function of the respective $%
\theta $-orders and $S_{..}$ acts on the respective tensor factor with the
remaining particle being a spectator. As before one checks that this vector
fulfills the modular localization equation $SJ_{0}\Delta ^{\frac{1}{2}}\psi
_{3}=\psi _{3};$ the Tomita operator acting on $\psi _{3}$ just re-shuffles
the six terms. As in the two-particle case, this action creates a vector
with a complicated incoming particle content having components to all
particle numbers. The last two terms correspond to nested mirror
permutations and, as will be seen below, results in the appearance of
``nondiagonal inclusive processes'' terms in the ($\psi _{3}^{\prime },\psi
_{3})$ inner product which generalize the diagonal inclusive processes \cite
{Essay} which result from the summation over final states in cross sections.

As an example we write down the integrand of one of those nondiagonal
inclusive terms 
\begin{equation}
\left\langle 0\left| a(\theta _{1}^{\prime })a(\theta _{2}^{\prime
})a(\theta _{3}^{\prime })S\cdot S_{12}^{\ast }a^{\ast }(\theta _{3})a^{\ast
}(\theta _{2})a^{\ast }(\theta _{1})\right| 0\right\rangle  \label{dot}
\end{equation}

In a graphical scattering representation particle 1 and 2 would scatter
first and produce arbitrarily many (subject to the conservation laws for the
total energy-momentum) particles which together with the third incoming
particle (which hitherto was only a spectator) enter an additional
scattering process of which only the 3-particle outgoing component is
separated out by the matrix element in (\ref{dot}). The dot means summation
over all admissable intermediate states and could be represented by e.g. a
heavy line in the graphical representation in order to distinguish it from
the one-particle lines. We will not write down the 6 contribution to the
inner product coming from the creation part and the remaining ones involving
the annihilation parts from the path $C.$

Our main interest is the study of $(\psi _{1}^{\prime },\psi _{3})$%
\begin{eqnarray}
&&(\psi _{1}^{\prime },\psi _{3}) \\
&=&\int_{C}\left\langle 0\left| a(\theta _{1}^{\prime })\{1\chi _{321}+S\chi
_{123}+S\cdot S_{12}\chi _{213}\}a(\theta _{3})a(\theta _{2})a(\theta
_{1})\right| 0\right\rangle \bar{f}_{1}^{\prime }(\theta _{1}^{\prime })\bar{%
f}_{3}(\theta _{3})\bar{f}_{2}(\theta _{2})\bar{f}_{1}(\theta _{1})  \notag
\\
&=&\int \left\langle 0\left| a(\theta _{1}^{\prime })\{1\chi _{321}+S+S\cdot
S_{12}\chi _{213}\}a(\theta _{3})a^{\ast }(\theta _{2})a^{\ast }(\theta
_{1})\right| 0\right\rangle \bar{f}_{1}^{\prime }(\theta _{1}^{\prime })\bar{%
f}_{3}(\theta _{3})f_{2}(\theta _{2})f_{1}(\theta _{1})  \notag \\
&=&\int \left\langle 0\left| a(\theta _{1}^{\prime })a(\theta
_{3})\{1+S\}a^{\ast }(\theta _{2})a^{\ast }(\theta _{1})\right|
0\right\rangle \bar{f}_{1}^{\prime }(\theta _{1}^{\prime })\bar{f}%
_{3}(\theta _{3})f_{2}(\theta _{2})f_{1}(\theta _{1})  \notag
\end{eqnarray}
The last line is (apart from a $a(\theta _{2})$-$a^{\ast }(\theta _{1})$
contraction term) the only nonvanishing contribution. Here the S-factor in
front of the $S_{12}$ has been transferred as $S^{\ast }$ onto the left hand
one-particle vector whereupon it acts as the identity. Renaming $%
f_{3}(\theta _{3})\rightarrow f^{\prime }(\theta _{2}^{\prime })$ we obtain
the result (\ref{2}). We now apply the KMS property for inner products of
modular subspaces 
\begin{eqnarray}
&&\left( \psi ^{\prime },\psi \right) =\left( \psi ,\Delta \psi ^{\prime
}\right) \\
&&\psi ,\psi ^{\prime }\in \mathcal{H}_{R}(W)\subset \mathcal{H}_{Fock} 
\notag
\end{eqnarray}
For the case at hand $\psi ^{\prime }=\psi _{1}^{\prime },\psi =\psi _{3}$
the particle interpretation of this KMS relation for modular vectors is
precisely the crossing symmetry relation. For the more general case
antiparticles$\neq $particles one has to work in the dense complex subspace $%
\mathcal{H}_{R}(W)+i\mathcal{H}_{R}(W)$ (which is a complete Hilbert space
in its own right in the thermal topology \cite{Sch1}. The conversion of the
KMS property for the inner product of the modular localized state vectors
with n-1 $f$ labels with a one-particles vector containing one $f$ contains
the crossing information for scattering of $n_{in}+n_{out}=n$ particles.

The crucial question is whether these inner products can also be used in
order to define correlation functions of PFG n-point functions 
\begin{equation}
\left\langle 0\left| F(f_{1})...F(f_{n})\right| 0\right\rangle
\end{equation}
for the lowest nontrivial case of 4-point function we already checked one
such condition (\ref{con}). I have carried out other consistency checks and
do not think that the prerequisites of \cite{BBS} leading back to particle
conservation can be derived from these correlation functions. The question
if and how these would be correlation functions are related to the
perturbative on-shell S-matrix representations mentioned in the introduction
is particularly interesting I hope to return to the issue of the form of
PFG-correlation functions in a more complete and systematic way in a future
paper.

For non-factorizing theories the interest in the modular localization
approach \ is (besides the improvement in the understanding the structure of
interacting QFT) the possible existence of an on-shell perturbation theory
of local nets avoiding the use of the non-intrinsic field coordinatizations.
This is a the revival of the perturbative version of the old dream to
construct S-matrices just using crossing symmetry in addition to unitarity
and no pointlike fields. The old S-matrix bootstrap program admittedly did
not get far, but now we perhaps can formulate a similar but structurally
richer problem as a perturbative approach to correlation functions of the
on-shell PFGs. Modular theory has given us a lot of insight and nobody
nowadays would try to cleanse the Einstein causality and locality concepts
from the stage as it was done in the 60$^{ies}.$ To the contrary, the local
off-shell observable algebras would be in the center of interest and the
avoidance of quantization would have entirely pragmatic reasons. In
particular the sharpening of localization beyond wedges is done by algebraic
intersections of wedge algebras rather than by cut-off or test function
manipulations on field coordinates.

The successful d=1+1 bootstrap-formfactor program of the previous section
for factorizing models yields S-matrices and formfactors which for models
with a continuous coupling are analytic around g=0. A good illustration is
the Sine-Gordon theory \cite{BFKZ}. The more local off-shell quantities
however (i.e. pointlike field operators or operators from algebras belonging
to bounded regions) are radically different since they involve virtual
particle polarization clouds which formally may be represented by infinite
series in the on shell Fs similar to the factorizing d=1+1 case of the
previous section. The analytic status of these quantities (i.e. localized
operators and their correlation functions) is presently not known; it may
well turn out that they are only Borel summable or (in the general
non-factorizable case) worse. The on-shell/off-shell dichotomy of the
modular approach for the first time allows to determine more precisely if
the cause of the possible breakdown of analyticity at g=0 are the
polarization clouds.

A solution of these problems, even if limited to some new kind of
perturbation theory (perturbation theory of wedge algebras and their
intersections) should also shed some light on the question of how to handle
theories involving higher spin particles, which in the standard off-shell
causal perturbation theory lead to short distance non-renormalizability. A
very good illustration of what I mean is the causal perturbation of massive
spin=1 vectormesons. Here the coupling of covariant fields obtained by
covariantizing the Wigner particle representation theory in the sense of the
previous section will not be renormalizable in the sense of short distance
power counting. In the standard perturbative approach the indefinite metric
ghosts are used to lower the operator dimension of the interaction densities
(free field polynomials) $W(x),$ which as a result of the free vectormeson
dimension dim$A_{\mu }=2,$ are at least 5, down to the value 4 permitted by
the renormalization requirements in a d=1+3 causal perturbative approach 
\cite{Du-Sch}. Since the ghosts are removed at the end, the situation is
akin to a catalyzer in chemistry: they do not appear in the original
question and are absent in the final result (without leaving any intrinsic
trace behind). In theoretical physics the presence of such catalyzers should
be understood as indicating that the theory wants to be analyzed on a deeper
level of local quantum physics i.e. further away from quantization and
quasiclassics. Indeed in the present on-shell modular approach the short
distance operator reason for introducing such ghosts would not be there and
the remaining question is again whether the modular program allows for a
perturbative analytically managable formulation.

\section{\protect\bigskip The AQFT Framework}

After our pedestrian presentation of the wedge algebra approach, it is time
to be more systematic and precise. For noninteracting free system the
conversion of the rather simple spatial nets of real subspaces of the Wigner
space of momentum space (m.,s) wave functions into a interaction-free net in
Fock space produces with the following three properties which continue to
hold in the presence of interactions. They have been explained in many
articles \cite{Bu} and in a textbook \cite{Ha}, and my main task here is to
adapt them to the problems of this paper.

\begin{enumerate}
\item  \textit{A net of local} (C$^{\ast }$- or von Neumann) \textit{%
operator algebras} indexed by classical spacetime regions $\mathcal{O}$%
\begin{equation*}
\mathcal{O}\rightarrow \mathcal{A}(\mathcal{O})
\end{equation*}
Without loss of generality the regions $\mathcal{O}$ maybe restricted to the
Poincar\'{e} covariant family of general double cones and the range of this
map may be described in terms of concrete operator algebras in Hilbert space
for which the vacuum representation $\pi _{0}$ may be taken i.e. $\mathcal{A}%
(\mathcal{O})\equiv \pi _{0}(\mathcal{A}(\mathcal{O})).$ The geometrical and
physical coherence properties as isotony: $\mathcal{A}(\mathcal{O}%
_{1})\subset \mathcal{A}(\mathcal{O}_{2})$ for $\mathcal{O}_{1}\subset 
\mathcal{O}_{2}$ and Einstein causality: $\mathcal{A}(\mathcal{O}^{^{\prime
}})\subset \mathcal{A}(\mathcal{O})^{\prime }$ are then evident coherence
requirements$.$ Here we use the standard notation of AQFT: the dash
superscript on the region denotes the causal disjoint and on the von Neumann
algebra it stands for the commutant within $B(H)$ where $H$ is the ambient
Hilbert space (here the representation space of the vacuum representation).
Einstein causality can be interpreted as an a priori knowledge about some
with $\mathcal{A}(\mathcal{O})$ commensurable observables in the sense of
von Neumann. This causality property suggests the question if complete
knowledge about commensurability $\mathcal{A}(\mathcal{O}^{\prime })=%
\mathcal{A}(\mathcal{O})^{\prime }$ is possible. It turns out that this is
indeed the generic behavior of vacuum nets called Haag duality. The cases of
violation of this duality are of particular interest since they can be
related to a very fundamental intrinsic characterization of spontaneous
symmetry breaking, thus vastly generalizing the Nambu-Goldstone mechanism
which was abstracted from quantization \cite{Ha}.

\item  \textit{Poincar\'{e} covariance and spectral properties}. 
\begin{eqnarray*}
g &\in &\mathcal{P}\rightarrow \alpha _{g}\,\,\,automorphism \\
&&\alpha _{g}(\mathcal{A}(\mathcal{O}))=\mathcal{A}(\mathcal{O})
\end{eqnarray*}
is unitarily implements in the vacuum representation 
\begin{eqnarray*}
U(g)AU^{\ast }(g) &=&\alpha _{g}(A) \\
A &\in &\mathcal{A}(\mathcal{O})
\end{eqnarray*}
The unitaries for the translations have energy-momentum generators which
fulfil the relativistic spectrum (positive energy) condition, symbolically $%
specU(a)\in \overline{V^{\upharpoonleft }}$ (the closed forward light cone)

\item  The phase space structure of local quantum physics or the
``nuclearity property''.
\end{enumerate}

\begin{remark}
The precise fomulation of the third property is somewhat involved and will
be presented after the following remarks on the first two structural
properties. Since in the formulation of the net one may work without loss of
generality with von Neumann algebras \cite{Ha}, the first question is what
type in the Murray-von Neumann-Connes-Haagerup classification occurs. There
is a very precise answer for wedges (which may be considered as double cones
at infinity). As a result of the existence of a one-sided translation into a
wedge as well as of the split property below, the wedge algebras $\mathcal{A}%
(W)$ turn out to be a hyperfinite factor of type III$_{1}.$ This implies in
particular that the algebra has properties which take it far away from the
structure of QM (factors of type I$_{\infty })$. Such algebras do not have
pure states or minimal projectors, rather all faithful states on such
algebras are thermal i.e. obey the KMS condition. This makes them similar to
states appearing in CST with bifurcated horizons as in Hawking-Unruh
situations however with modular flows instead of Killing flows.(but more
``quantum''.i.e. without the classical geometric Killing vector aspects of
horizons). The modular flow near the boundary of e.g. double cone regions
become asymptotically geometric and Killing-like.
\end{remark}

The nuclearity requirement results from the idea that there should be a
local quantum physical counterpart of the phase space properties of QM in a
box. The famous \textit{finite number of degrees of freedom law per unit
cell of QM phase space} results from limiting the discrete box spectrum by a
cut-off in energy. As first suggested by Haag and Swieca \cite{Ha}, the
corresponding LQP counterpart, based on the causally closed double cone
analogue of the quantization box in Schr\"{o}dinger QM, points into the
direction of a ``weakly'' infinite number; according to their estimates this
set of state vectors was compact in Hilbert space. Subsequent refinements of
techniques revealed that this set is slightly smaller namely ``nuclear'' 
\cite{Ha}, and exact calculations with interaction-free theories
demonstrated that the phase space situation also cannot be better than
nuclear.

The best way to understand this issue is to follow the motivating footsteps
of Haag and Swieca. They, as many other physicists at that time (and as some
contemporary philosophers \cite{Cl-Hal}), were attracted by the intriguing
consequences of the of the so-called Reeh-Schlieder property of QFT 
\begin{eqnarray}
\overline{\mathcal{P}(\mathcal{O})\Omega } &=&H,\,\,cyclicity\,\,of\,\,\Omega
\label{cyc} \\
A\in \mathcal{P}(\mathcal{O}),\,\,A\Omega &=&0\Longrightarrow
A=0\,\,i.e.\,\,\Omega \,\,separating  \notag
\end{eqnarray}
which either holds for the polynomial algebras of fields (which are
affiliated to the von Neumann algebras which they generate) or for operator
algebras $\mathcal{A}(\mathcal{O}).$ The first property, namely the
denseness of states created from the vacuum by operators from arbitrarily
small localization regions (e.g. a state describing ``a particle behind the
moon'\footnote{%
This weird aspect should not be held against QFT but rather be taken as
indicating that localization by a piece of hardware in a laboratory is also
limited by an arbitrary large but finite energy, i.e. is a ``phase space
localization'' (see subsequent discussion). In QM one obtains genuine
localized subspaces without energy limitations.} and a charge compensating
antiparticle in some other far away region can be approximated inside a
laboratory of arbitrary small size and duration) is totally unexpected from
the global viewpoint of general QT. In the algebraic formulation this can be
shown to be dual to the second one (in the sense of passing to the
commutant), in which case the cyclicity passes to the separating property of 
$\Omega $ with respect to $\mathcal{A}(\mathcal{O}^{\prime }).$ Referring to
its use, the separating property is often called the \textit{field}- \textit{%
state relation}. The mathematical terminology is to say that the pair ($\ 
\mathcal{A}(\mathcal{O})$,$\Omega )$ is ``standard''. The large enough
commutant required by the latter property is guarantied by causality (the
existence of a nontrivial causal disjoint $\mathcal{O}^{\prime })$ and thus
shows that causality is again responsible for this unexpected denseness
property.

Of course the claim that somebody causally separated from us may provide us
nevertheless with a dense set of states is somewhat perplexing especially if
one compares it with the tensor factorization properties of good old
Schr\"{o}dinger QM with respect to an inside/outside separation via a
quantization box.

If the naive interpretation of cyclicity/separability in the Reeh-Schlieder
theorem leaves us with a feeling of science fiction (and for this reason as
already mentioned justifiably also has attracted attention in philosophical
quarters), the challenge for a theoretical physicist is to find an argument
why, for all practical purposes, the situation nevertheless remains similar
to QM. This amounts to the fruitful question which vectors among the dense
set of state vectors can be really produced with a controllable expenditure
(of energy); a problem from which Haag and Swieca started their
investigation. In QM this question was not that interesting, since the
localization at a given time via support properties of wave functions leads
to a tensor product factorization of inside/outside so that the inside state
vectors are evidently never dense in the whole space and the ``particle
behind the moon paradox'' does not occur.

Later we will see that most of the very important physical and geometrical
informations are encoded into features of dense domains, in fact the
aforementioned modular theory is explaining this deep relation between
operator domains of the Tomita $S$ and spacetime geometry. As mentioned
before the individuality of the various S-operators is only the difference
in domains, since all of them act as $SA\Omega =A^{\ast }\Omega ,A\in 
\mathcal{A}(\mathcal{O})$

For the case at hand the reconciliation of the paradoxical aspect \cite{JPA}
of the Reeh-Schlieder theorem with common sense has led to the discovery of
the physical relevance of \textit{localization with respect to phase space
in LQP}, i.e. the understanding of the \textit{size of degrees of freedom}
in the set: (notation $\mathbf{H}=\int EdP_{E})\,\,$ 
\begin{eqnarray}
&P_{E}\mathcal{A}(\mathcal{O})\Omega \,\,is\;compact&\,\,\, \\
&P_{E}\mathcal{A}(\mathcal{O})\Omega \,\,or\,\,e^{-\beta \mathbf{H}}\mathcal{%
A}(\mathcal{O})\Omega \;is\,\,nuclear&
\end{eqnarray}
The first property was introduced way back by Haag and Swieca \cite{Ha}
whereas the second more refined statement (and similar nuclearity statements
involving modular operators of local regions instead of the global
hamiltonian) which is saturated by QFT (i.e. cannot be improved) and easier
to be used, is a later result of Buchholz and Wichmann \cite{Bu-Wi}. It
should be emphasized that the LQP degrees of freedom counting of
Haag-Swieca, which gives an infinite but still compact (and even nuclear)
set of phase-space localized states, is different from the QM finiteness of
degrees of freedom per phase used in some contemporary entropy calculations.

The map $\mathcal{A}(\mathcal{O})\rightarrow e^{-\beta H}\mathcal{A}(%
\mathcal{O})\Omega $ is \ only nuclear if the mass spectrum of LQP is not
too accumulative in finite mass intervals; in particular infinite towers of
equal mass particles are excluded (which then would cause the strange
appearance of a maximal ``Hagedorn'' temperature). The nuclearity assures
that a QFT, which was given in terms of its vacuum representation, also
exists in a thermal state. An associated nuclearity index turns out to be
the counterpart of the quantum mechanical Gibbs partition function \cite{Bu} 
\cite{Ha} and behaves in an entirely analogous way.

The peculiarities of the above degrees-of freedom-counting are very much
related to one of the oldest ``exotic'' and at the same time characteristic
aspects of QFT, namely vacuum polarization. As first observed by Heisenberg,
the partial charge: 
\begin{equation}
Q_{V}=\int_{V}j_{0}(x)d^{3}x=\infty  \label{div}
\end{equation}
diverges as a result of uncontrolled vacuum particle/antiparticle
fluctuations near the boundary. For the free field current it is easy to see
that a better definition involving test functions, which smoothens the
behavior near the boundary and takes into account the fact that the current
is a 4-dim. distribution which has no restriction to equal times, leads to a
finite expression.

\begin{equation}
Q_{R}=\int j_{0}(x)f(x_{0})g(\frac{\mathbf{x}}{R})d^{s}x
\end{equation}
where f and g are test functions of compact support with $\int
f(x_{0})dx_{0}=1$ and $g(\vec{x})=1$ for $\left| \vec{x}\right| <1$ and $g(%
\vec{x})=0$ $\left| \vec{x}\right| >1+\delta .$ The vectors $Q_{R}\Omega $
only converge weakly for $R\rightarrow \infty $ on a dense domain. Their
norms however diverge as \cite{BDLR} 
\begin{align}
\left( Q_{R}\Omega ,Q_{R}\Omega \right) & \leq const\cdot R^{s-1}
\label{surface} \\
& \sim area  \notag
\end{align}
The surface-layer character of this vacuum polarization is reflected in this
area behavior together with the original divergence (\ref{div}) for fixed $R$
and $\delta \rightarrow 0$.

The algebraic counterpart is the so called ``split property'', namely the
statement \cite{Ha} that if one leaves between say the double cone (the
inside of a ``relativistic box'') observable algebra $\mathcal{A}(\mathcal{O}%
)\,$and its causal disjoint (its relativistic outside) $\mathcal{A}(\mathcal{%
O}^{\prime })$ a ``collar'' (geometrical picture of the relative commutant) $%
\mathcal{O}_{1}^{\prime }\cap \mathcal{O}$, i.e. 
\begin{equation}
\mathcal{A}(\mathcal{O})\subset \mathcal{A}(\mathcal{O}_{1}),\,\,\,\mathcal{%
O\ll O}_{1}\,,\,\,properly
\end{equation}
then it is possible to construct in a canonical way a type $I$ tensor factor 
$\mathcal{N}$ which extends in a ``fuzzy'' manner into the collar $\mathcal{A%
}(\mathcal{O})^{\prime }\cap \mathcal{A}(\mathcal{O}_{1})$ i.e. $\mathcal{A}(%
\mathcal{O})\subset \mathcal{N}\subset \mathcal{A}(\mathcal{O}_{1}).$ With
respect to $\mathcal{N}$ the Hilbert space factorizes i.e. as in QM; there
are states with no fluctuations (or no entanglement) for the ``smoothened''
operators in $\mathcal{N}.$ Whereas the original vacuum will be entangled
from the box point of view, there also exists a disentangled product vacuum
on $\mathcal{N}.$ The algebraic analogue of a smoothening of the boundary by
a test function is the construction of a factorization of the vacuum with
respect to a suitably constructed type $I$ factor algebra which uses the
above collar extension of $\mathcal{A}(\mathcal{O}).$ It turns out that
there is a canonical, i.e. mathematically distinguished factorization, which
lends itself to define a natural ``localizing map'' $\Phi $ and which has
given valuable insight into an intrinsic LQP version of Noether's theorem 
\cite{Ha}, i.e. one which does not rely on quantizing classical Noether
currents. It is this ``split inclusion'' which allows to bring back the
familiar structure of pure states, tensor product factorization,
entanglement and all the other properties at the heart of standard quantum
theory and the measurement process. However despite all the efforts to
return to structures known from QM, the original vacuum retains its thermal
(entanglement) properties with respect to all localized algebras, even with
respect to the ``fuzzy''-localized $\mathcal{N}.$

Let us collect in the following some useful mathematical definitions and
formulas for ``standard split inclusions'' \cite{Do-Lo}

\begin{definition}
An inclusion $\Lambda =(\mathcal{A},\mathcal{B},\Omega )$ of factors is
called standard split if the collar $\mathcal{A}\prime \cap \mathcal{B}$ as
well as $\mathcal{A},\mathcal{B}$ together with $\Omega $ are standard in
the previous sense, and if in addition it is possible to place a type I$%
_{\infty }$ factor $\mathcal{N}$ between $\mathcal{A}$ and $\mathcal{B}$.
\end{definition}

In this situation there exists a canonical isomorphism of $\mathcal{A}\vee 
\mathcal{B}^{\prime }$ to the tensor product $\mathcal{A}\bar{\otimes}%
\mathcal{B}^{\prime }$ which is implemented by a unitary $U(\Lambda
):H_{\Lambda }\rightarrow H_{1}\bar{\otimes}H_{2}$ (the ``localizing map'')
with 
\begin{eqnarray}
&&U(\Lambda )(AB^{\prime })U^{\ast }(\Lambda )=A\bar{\otimes}B^{\prime } \\
&&A\in \mathcal{A},\,\,B^{\prime }\in \mathcal{B}^{\prime }  \notag
\end{eqnarray}
This map permits to define a canonical intermediate type I factor $\mathcal{N%
}_{\Lambda }$ (which may differ from the $\mathcal{N}$ in the definition) 
\begin{equation}
\mathcal{N}_{\Lambda }:=U^{\ast }(\Lambda )(B(H_{1})\otimes \mathbf{1)}%
U(\Lambda )\subset \mathcal{B}\subset B(H_{\Lambda })
\end{equation}
It is possible to give an explicit formula for this canonical intermediate
algebra in terms of the modular conjugation $J=U^{\ast }(\Lambda )J_{%
\mathcal{A}}\otimes J_{\mathcal{B}}U(\Lambda )\,$\ of the collar algebra ($%
\mathcal{A}^{\prime }\cap \mathcal{B},\Omega )$ \cite{Do-Lo} 
\begin{equation}
\mathcal{N}_{\Lambda }=\mathcal{A}\vee J\mathcal{A}J=\mathcal{B}\wedge J%
\mathcal{B}J
\end{equation}
\thinspace 

The tensor product representation gives the following equivalent tensor
product representation formulae for the various algebras 
\begin{eqnarray}
\mathcal{A} &\sim &\mathcal{A}\otimes \mathbf{1} \\
\mathcal{B}^{\prime } &\sim &\mathbf{1}\otimes \mathcal{B}^{\prime }  \notag
\\
N_{\Lambda } &\sim &B(H_{\Lambda })\otimes \mathbf{1}  \notag
\end{eqnarray}
As explained in \cite{Do-Lo}, the uniqueness of $U(\Lambda )$ and $%
N_{\Lambda }$ is achieved with the help of the ``natural cones'' $\mathcal{P}%
_{\Omega }(\mathcal{A}\vee \mathcal{B}^{\prime })$ and $\mathcal{P}_{\Omega
\otimes \Omega }(\mathcal{A}\otimes \mathcal{B}^{\prime }).$ These are cones
in Hilbert space whose position in $H_{\Lambda }$ together with their facial
subcone structures preempt the full algebra structure on a spatial level.
The corresponding marvelous theorem of Connes \cite{Connes} goes far beyond
the previously mentioned state vector/field relation.

Returning to our physical problem, we have succeeded to find the right
analogue of the QM box. Contrary to the causally closed local type III
algebras with their sharp light cone boundaries (``quantum horizons''), the
``fuzzy box'' type I factor $N_{\Lambda }$ permits all the structures we
know from QM: pure states, inside/outside tensor factorization,
(dis)entanglement etc. with one exception: the vacuum is highly entangled in
the tensor product description; the modular group of \ the state $\omega
\mid _{\mathcal{A}\bar{\otimes}\mathcal{B}^{\prime }}$represented in the
tensor product natural cone $P_{\Omega \otimes \Omega }(\mathcal{A}\bar{%
\otimes}\mathcal{B}^{\prime })$ is not the tensorproduct of the modular
groups of $\mathcal{A}$ and $\mathcal{B}^{\prime }$, whereas the modular
conjugation $J$ acts on the tensor product cone as $J_{\mathcal{A}}\bar{%
\otimes}J_{\mathcal{B}}$ (since the restriction $\omega \mid _{\mathcal{A%
\bar{\otimes}B}^{\prime }}$is faithful)$.$ Note also that the restriction of
the product state $\omega \otimes \omega $ to $\mathcal{B}$ or $\mathcal{B}%
^{\prime }$ is not faithful resp. cyclic on the corresponding vectors and
therefore the application of those algebras to the representative vectors $%
\eta _{\omega \otimes \omega }$ yields nontrivial projectors (e.g. $%
P_{\Lambda }=U^{\ast }(\Lambda )B(H_{1})\bar{\otimes}1U(\Lambda )).$

Since the fuzzy box algebra $N_{\Lambda }$ is of quantum mechanical type I,
we are allowed to use the usual trace formalism based on the density matrix
description, i.e. the vacuum state can be written as a density matrix $\rho
_{\Omega }$ on $\mathcal{N}_{\Lambda }$ which leads to a well-defined von
Neumann entropy 
\begin{eqnarray}
\left( \Omega ,A\Omega \right) &=&tr\rho _{\Lambda }A,\,\,A\in \mathcal{A} \\
S(\rho _{\Lambda }) &=&-tr\rho _{\Lambda }log\rho _{\Lambda }
\end{eqnarray}
but this is not sufficient to determine $\rho _{\Lambda }$ which is needed
for the von Neumann entropy of the fuzzy box $S(\rho _{\Lambda }).$ If we
would be able to compute the unitary representer $\Delta _{\mathcal{N}%
_{\Lambda }}^{it}$ of the modular group of the pair ($\mathcal{N}_{\Lambda
},\Omega )$ then we know also $\rho _{\Lambda }$ since the modular operator
of a type I factor is known to be related to an unnormalized density matrix $%
\check{\rho}_{\Lambda }$ with $\rho _{\Lambda }=\frac{1}{tr\check{\rho}%
_{\Lambda }}\check{\rho}_{\Lambda }$ through the tensor product formula on $%
H_{1}\bar{\otimes}H_{2}$ $\ $%
\begin{equation}
\Delta =\check{\rho}_{\Lambda }\bar{\otimes}\check{\rho}_{\Lambda }^{-1}
\end{equation}
Actually there are several intuitively equivalent definitions of
localization entropy \cite{Narn} Among those the most convenient one seems
to be the relative entropy of the vacuum $\omega $ with respect to the split
vacuum $\omega \times \omega .$ The relative entropy of a von Neumann
algebra $\mathcal{M}$ of one faithful state $\omega _{1}$ with respect to $%
\omega _{2}$ uses the relative modular operator \cite{Ha} $\Delta _{\omega
_{1},\omega _{2}}$

\begin{equation}
S(\omega _{1}|\omega _{2})_{M}=-\left\langle log\Delta _{\omega _{1},\omega
_{2}}\right\rangle
\end{equation}
Kosaki \cite{Kosaki} was able to convert this (in the most general setting)
into a variational formula 
\begin{align}
S(\omega _{1}|\omega _{2})_{\mathcal{M}}& =sup\int_{0}^{1}\left[ \frac{%
\omega (1)}{1+t}-\omega _{1}(y^{\ast }(t)y(t))-\frac{1}{t}\omega
_{2}(x^{\ast }(t)x(t))\frac{dt}{t}\right] \\
x(t)& =1-y(t),\,\,x(t)\in \mathcal{M}  \notag
\end{align}
where in our case $\omega _{1}=\omega \times \omega ,\,\,\omega _{2}=\omega
,\,\,\mathcal{M}=\mathcal{A}\vee \mathcal{B}^{\prime }.$

Despite the very clear conceptual setting of this split entropy, it is
difficult to obtain good estimates for this entropy, not to mention exact
calculations. As for the above partial charges (\ref{surface}) one expects a
surface behavior, the quantum version of the Bekenstein-Hawking area law. An
existing estimate shows that its increase for e.g. double cones is weaker
than the spatial volume \cite{Narn}. The most accessible situation for
entropy calculations seems to be conformal QFT

It seems that for double cones in conformal theories one cane use the
geometric aspects of the situation and do an explicit calculation. This
still needs to be carried out, but an outline of the strategy can be found
in \cite{Essay}.

Ideas about localization entropy are quite inaccessible in perturbation
theory because they require an intrinsic description in terms of a net of
algebras, whereas for perturbation theory no description without the use of
field-coordinatizations is known. This has led to speculative remarks in the
literature claiming the necessity of new degrees of freedom for the
understanding of the area law.

\section{\protect\bigskip Modular Inclusions and Intersections, Holography}

One of the oldest alternative proposals for canonical (equal time)
quantizations is the so called light ray or light front (or $p\rightarrow
\infty $ frame) quantization. The trouble with it is that it apparently
inherits some the short distance diseases from the canonical quantization.
The latter is known to only makes sense for superrenormalizable interactions
but not for strictly renormalizable ones, which lead to infinite
multiplicative renormalizations. Let us ignore this for a moment and look at
some additional problems of light cone quantization which canonical equal
time quantization does not have. This is the apparent loss of the connection
with local QFT; in fact in none of the papers on light cone quantization it
is spelled out how to return to a local QFT. The problem of light front
restricted free fields was rigorously studied in \cite{Dim}, but in the
interacting case the reconstruction of the local theory from that on the
light cone (which may be called the holographic reconstruction) is a serious
problem indeed.

Our modular inclusion techniques in section 2 suggested that for massive
(and massless for $d\neq 1+1$) theories the wedge algebra and the chiral
light front algebra are identical 
\begin{equation}
A(W)=A(\mathbb{R}_{>})
\end{equation}
Since this is a consistent property which is fulfilled by all known quantum
field theoretic models, we will be focus our interests on theories which
obey this ``characteristic shadow'' property and leave open the question
whether this property is a consequence of the standard physical requirements
of AQFT. We already mentioned in the same section that the chiral algebra
really should be thought of as the ``transversally unresolved light front
algebra''. But since the use of a light front notation like $A(\mathbb{R}%
_{>}^{d-1})$ could suggest the wrong idea that one deals with a full light
front net, we prefer the light ray notation since it does not count any
localizationwise unresolved dimensions. If we just refer to the global
algebras and not to their local (sub)net structure, then all three objects
are equal and there could be no confusion.

The rigorous construction of a chiral net for $\mathcal{A}(\mathbb{R}_{>})$
indicated in the third section will now be presented in more detail within
its natural setting of modular inclusions \cite{GLW}.

One first defines an abstract modular inclusion in the setting of von
Neumann algebras. There are several types of inclusions which have received
mathematical attention\footnote{%
In addition to the split inclusion used in the previous section, there are
the famous V. Jones inclusions whose charateristic property is the existence
of conditional expectations. Their domain in particle physics is in the area
of charge fusion and internal symmetry.}. An inclusion of two factors $%
\mathcal{N}\subset \mathcal{M}$ is called (+ halfsided) modular if the
modular group $\Delta _{\mathcal{M}}^{it}$ for $t<0$ transforms $\mathcal{N}$
into itself (compression of $\mathcal{N}$) 
\begin{equation}
Ad\Delta _{\mathcal{M}}^{it}\mathcal{N}\subset \mathcal{N}
\end{equation}
We assume that $\cup _{t}Ad\Delta _{\mathcal{M}}^{it}\mathcal{N}$ is dense
in $\mathcal{M}$ (or that $\cap _{t}\Delta _{\mathcal{M}}^{it}\mathcal{N=}%
\mathbb{C\cdot }1)$. This means in particular that the two modular groups $%
\Delta _{\mathcal{M}}^{it}$ and $\Delta _{\mathcal{N}}^{it}$ generate a two
parametric group of $\left( \text{translations, dilations}\right) $ in which
the translations have positive energy \cite{Wies}. Let us now look at the
relative commutant as done e.g. in the appendix of \cite{S-W3}.

Let $(\mathcal{N\subset M},\Omega )$ be modular with nontrivial relative
commutant. Then look at the subspace generated by relative commutant $%
H_{red}\equiv \overline{(\mathcal{N}^{\prime }\cap \mathcal{M)}\Omega }%
\subset H.$ The modular groups to $\mathcal{N}$ and $\mathcal{M}$ leave
invariant this subspace: $\Delta _{\mathcal{M}}^{it},t<0$ maps $\mathcal{N}%
^{\prime }\cap \mathcal{M}$ into itself by the inclusion being modular. Look
at the orthogonal complement of $H_{red}$ in $H.$ This orthogonal complement
is mapped into itself by $\Delta _{\mathcal{M}}^{it}$ for positive $t.$ Let $%
\psi $ be in that subspace, then 
\begin{equation}
\left\langle \psi ,\Delta _{\mathcal{M}}^{it}(\mathcal{N}^{\prime }\cap 
\mathcal{M})\Omega \right\rangle =0\,\,for\,\,t>0.
\end{equation}

$\,$Analyticity in $t$ then gives the vanishing for all $t.$

Due to Takesaki's theorem \cite{Tak}, we can restrict $\mathcal{M}$ to $%
H_{red}$ using a conditional expectation to this subspace defined in terms
of the projector $P$ onto $H_{red}$. Then 
\begin{eqnarray}
E(\mathcal{N}^{\prime }\cap \mathcal{M)} &\subset &\mathcal{M}|_{\overline{(%
\mathcal{N}^{\prime }\cap \mathcal{M)}\Omega }}=E(\mathcal{M}) \\
E(\cdot ) &=&P\cdot P
\end{eqnarray}
is a modular inclusion on the subspace $H_{red}.$ $\mathcal{N}$ also
restricts to that subspace and this restriction is obviously in the relative
commutant of $E(\mathcal{N}^{\prime }\cap \mathcal{M)\subset }E(\mathcal{M)}$%
. Moreover using arguments as above it is easy to see that the restriction
is cyclic with respect to$\,\Omega $ on this subspace. Therefore we arrive
at a reduced modular ``standard inclusion'' 
\begin{equation}
(E(\mathcal{N)}\subset E(\mathcal{M}),\Omega )
\end{equation}
Standard modular inclusions are known to be isomorphic to chiral conformal
field theories \cite{GLW}.

This theorem and its extension to modular intersections leads to a wealth of
physical applications in QFT, in particular in connection with ``hidden
symmetries'' symmetries which are of purely modular origin and have no
interpretation in terms of quantized Noether currents \cite{S-W1}\cite{S-W3}%
. The modular techniques unravel new structures which are not visible in
terms of field coordinatizations. Holography and problems of degrees of
freedom counting (phase space in LQP) as well as the issue of localization
entropy are other examples.

Let us briefly return to applications for d=1+1 massive theories. It is
clear that in this case we should use the two modular inclusions which are
obtained by sliding the (right hand) wedge into itself along the upper/lower
light ray horizon. Hence we chose $\mathcal{M=A(}W\mathcal{)}$ and $\mathcal{%
N}=\mathcal{A}(W_{a_{+}})$ or $\mathcal{N}=\mathcal{A}(W_{a_{-}})$ where $%
W_{a\pm }$ denote the two upper/lower light like translated wedges $W_{a\pm
}\subset W.$ As explained in section 2 following (\cite{GLRV}) and mentioned
above, we do not expect the appearance of a nontrivial subspace (i.e. we
expect $P=1)$ in the action of the relative commutants onto the vacuum 
\begin{eqnarray}
&&\mathcal{A}(I(0,a_{\pm }))\equiv A(W_{a_{\pm }})^{\prime }\cap \mathcal{A}%
(W) \\
&&\overline{\mathcal{A}(I(0,a_{\pm })\Omega }=H  \notag
\end{eqnarray}
where the notation indicates that the localization of $\mathcal{A}%
(I(0,a_{\pm }))$ is thought of as the piece of the upper/lower light ray
interval between the origin and the endpoint $a_{\pm }$.

From the standardness of the inclusion one obtains according to the previous
discussion an associated conformal net on the line, with the following
formula for the chiral conformal algebra on the half line 
\begin{equation}
\mathcal{A}_{\pm }(R_{>})\equiv \bigcup_{t\geq 0}Ad\Delta _{W}^{it}\left( 
\mathcal{A}(I(0,a_{\pm }))\right) \subseteq \mathcal{A}(W)\text{,}
\end{equation}
We expect the equality sign to hold 
\begin{equation}
\mathcal{A}_{\pm }(R_{>})=\mathcal{A}(W)
\end{equation}
but our argument was tied to the existence of PFGs since as a result of
their mass-shell structure 
\begin{equation}
F(\hat{f})=\int Z(\theta )f(\theta )d\theta =F_{res}(\hat{f}_{res})
\end{equation}
where the notation $res$ indicates the corresponding generators in light ray
theory which are identical in rapidity space and only differ in their
x-space appearance. This is a significant strengthening of the cyclicity
property $\overline{\mathcal{A}_{\pm }(R_{>})\Omega }=\overline{\mathcal{A}%
(W)\Omega }$ for the characteristic data on one light ray. The argument is
word for word the same in higher spacetime dimensions, since the appearance
of transversal components (which have no influence on the localization) in
addition to $\theta $ do not modify the argument. One would think that the
inference of PFG generators can even be disposed of and the equality should
follow from the standard causal shadow property of QFT in the form 
\begin{equation}
\mathcal{A}(W)=A(R_{>}^{(\alpha )})  \label{char}
\end{equation}
where $R_{>}^{(\alpha )}$ is a spacelike positive halfline with inclination $%
\alpha $ with respect to the x-axis. The idea is that if this relation would
remain continuous for $R_{>}^{(\alpha )}$ approaching the light ray ($\alpha
=45\unit{%
{{}^\circ}%
})$ which then leads to the desired equality. We believe that the relation (%
\ref{char}) for massive theories, which will be called ``characteristic
shadow property'', is a general consequence of the standard causal shadow
property (the identity of $\mathcal{A}(\mathcal{O})=\mathcal{A(O}^{\prime
\prime })$ where $\mathcal{O}^{\prime \prime }$ is the causal completion of
the convex spacelike region $\mathcal{O}$) in any spacetime dimension.

Theories with the characteristic shadow property are the objects of the
light ray folklore. The present conceptually more concise approach explains
why the light ray quantization in the presence of interactions is basically
nonlocal which significantly restricts its unqualified physical use. The
reason is that although the halfline algebra is equal to the wedge algebra
(since all rays of forward light cone propagation which pass through the
upper/lower half light ray $R_{>}$ have passes or will pass through $W)$,
the locality on the light ray cannot be propagated into the wedge (the
strips inside the wedge subtended from an interval $I$ on the light ray by
the action of the opposite light ray translation are for massive theories
not outside the propagation region of the complement of $I)$. Only for the
halfline itself one obtains a 2-dimensional shadow region namely the wedge
region. If one uses both light cones then it is possible to reconstruct a
causal d=1+1 net by intersections. This construction uses the
two-dimensional translation group on the wedge and the ensuing double cone
relative commutants. Note that in order to achieve this with parity
reflected halflines of light rays, one needs the relative position of the
two halfline light ray algebras relative to each other in the common space $%
H.$ In fact one shifted right light ray chiral algebra together with its
parity reflected image is equal to the union of two opposite spacelike
separated wedge algebras. The reflected light ray algebra may also be
replaced by the algebra on the left hand extension of the original light ray
since both create the same left wedge algebra. However the natural net
structure of that algebra is very nonlocal with respect to that of the
parity reflected one. This prevents its use in the construction of the
2-dim. net from shifts and geometric intersections on one light ray. An
algebra localized in an interval on one light ray corresponds to a
completely spread out algebra on the other ray. The modular group of each
light ray interval is geometric. This agrees with the qualitative behavior
one expects for the modular group of the double cone in a massive theory 
\cite{S-W1} near the causal horizon. Note that the relative nonlocality of
the chiral conformal theories is also necessary in order to be consistent
with a massive situation. The chiral conformal field theory contains the
standard light ray translation with a gapless spectrum. However this
spectrum is not the physical one since in that chiral theory there exists
yet another nonlocally acting translation and it is the spectrum of the
product of the two generators $P_{+}P_{-}$ which gives the physical mass.
Hence chiral conformal theories constitute a multipurpose tool in LQP. This
is why they can serve as ``holographic'' pieces for the construction of
massive d=1+1 theories. So with just one light ray and two translations, one
acting locally and the other nonlocally, one ray one can already reconstruct
the full d=1+1 net. Later we will see that this is enough to understand the
localization entropy which turns out to have the surface behavior first
observed in the context of classical localization behind black hole horizons
by Bekenstein.

Because of the transversal extension, the holography in terms of
one-dimensional chiral conformal theories is more complicated for higher
dimension. There one needs a family of chiral conformal theories which is
obtained from ``modular intersections''. Rather than associating the chiral
conformal theory with a light ray, it is more appropriate to associate it
with the transverse space of the wedge which contains the light ray i.e.
with the light front. A family is of light front algebras is obtained by
applying L-boosts to the standard wedge $W$ which tilts $W$ around one of
its defining light rays, so that the transversal degeneracy of the modular
inclusion is partially destroyed (in d=1+2 it would be completely
destroyed). In this way one obtains a fan-like ordered family of wedges
corresponding to a family of chiral conformal theories whose relative
position within the original Hilbert space contains all the informations
which are necessary in order to reconstruct the original (massive) theory. A
detailed and rigorous account of this construction will be given in a future
paper. Here we will only mention some analogies to the above light ray
situation. The process of tilting by applying a family of boost
transformations which leave the common light ray invariant is described by
unitary transformations of one chiral conformal theory into another. Each
single one, according to the higher dimensional characteristic shadow
property, is equal to a wedge algebra. Knowing the position of a finite
number of such chiral conformal theories with respect to each other (the
number increases with increasing spacetime dimensions), determines the
relative position of a finite number of wedge algebras ($\simeq $ chiral
conformal QFTs) which accoding to the previous remarks is sufficient to
reconstruct the original net (the blow-up property in \cite{S-W3}). As
previously mentioned, in the d=1+1 case the second light ray can be thought
of as obtained from the first one by a unitary parity reflection (assuming
that the theory is parity invariant). All the finitely many chiral conformal
field theories are unitarily equivalent (either by parity or by L-boosts);
the important physical information is contained in their relative position
within the same Hilbert space. The terminology ``scanning by a finite family
of chiral conformal theories'' is perhaps more appropriate for this
construction of higher dimensional theories \cite{S-W2}\cite{S-W3}.

It has been shown elsewhere \cite{S-W2} that the modular inclusion for two
wedges gives rise to two reflected eight-parametric subgroups of the
10-parametric Poincar\'{e} group which contain a two parametric transversal
Galilean subgroup of the type found by formal light front quantization
arguments \cite{Su}. All these considerations show the primordial role of
the chiral conformal QFT as a building block for the higher dimensional QFTs.

There is another much more special kind of holography in which an
isomorphism of a massive QFT in d+1 dimensions to a conformal d-dimensional
theory is in the focus of interest. This isomorphism appears in Rehren's
solution \cite{Re} of Maldacena's conjecture about the existence of a
holographic relation of quantum matter in a (d+1)-dimensional Anti de Sitter
spacetime with that in a d-dimensional conformal QFT. This type of
holography has not been observed outside the anti de Sitter spacetime and
since it is an isomorphism to a conformal theory, the degrees of freedom are
not really reduced in the sense of 't Hooft' \cite{Ho}, as it was the case
in the previous holography via light ray reduction. The Maldacena-Witten
holography is apparently of importance within the development of string
theory, in fact the protagonists believe that it contains information about
a possible message about quantum gravity of string theory. Within the
present AQFT setting its main interest is that it requires the
field-coordinatization free point of view in its strongest form: whereas in
most problems of QFT there exist appropriate field coordinatizations which
often facilitate calculations, the M-W isomorphism defined in rigorous terms
by Rehren is not pointlike and has no description in terms of fields outside
its algebraic version. In contradistinction to the light ray holography
which happens at the causality horizon (light front boundary) of modular
localization (or its classical Killing counterpart in case of black holes)
the AdS holography takes place at the boundary at infinity.

A very simple presentation in the spirit of Rehren's approach which takes
into account the covering of the relevant spaces can be found in \cite{Sch2}.

\section{Comparison with String Theory}

As mentioned in the introduction, historically string theory originated from
the attempt to understand and implement the issue of crossing symmetry of
the S-matrix. Without the intervention of QFT it was difficult to combine
unitarity and crossing symmetry into a manageable formalism. It came as
somewhat of a surprise that by assuming an additional stronger form of
crossing called ''duality'' one actually could obtain the dual model
formalism. Duality was an idea of entirely phenomenological origin which
consisted in the hypothesis that crossing can already hold if one only
restricts ones attention to (reggeized) one-particle states (``particle
democracy''). There was no theoretical support from QFT, nevertheless the
very appealing form of duality by Veneziano led eventually into string
theory. But whereas the content of QFT can be separated from the
perturbative formalism and cast into totally intrinsic form which is stronly
related with its underlying principles, string theory leaves a lot to be
desired on conceptual aspects and remained a collection of prescriptions. In
particular string theorists have not been able to successfully address the
issue of locality of operators and localization of states which are
absolutely crucial propertie on which any particle physics theory stands and
falls and which are even indispensable for the physical interpretation of
its formalism \cite{JPA}. The formal basis of string theory is a kind of
momentum space ``engineering'' rather than a conceptual spacetime analysis.
The latter remained within the realm of quasiclassical physics using
geometrical pictures with some fluctuation caused fuzziness i.e. pictures
which in the setting of quantum theory fall behind Heisenbergs dictum that
positions and momenta are not properties of the electron but are
characeristica of the events involving interactions with a measurement
apparatus which causes the factualization of potentialities. Related to this
is the fact that the word scattering theory has an entirely different
meaning in both areas. Whereas in QFT it is an asymptotic relation to free
fields for whose derivation spacelike locality is absolutely essential, in
string theory its use in the sense of the $0/\infty $ behavior of the
analytically continued source space conformal field theory in the complex
plane has nothing to do with any standard scattering concept of physical
particles. Whereas all important ideas in QFT have been tested outside
quasiclassical or perturbative settings at least in d=1+1 interacting
theories, this is not the case in string theory. For example the
Klein-Kaluza mechanism for the conversion of spacetime into inner symmetry
which is a (semi)classical idea has never been tested in a full QFT. Since
the physical origin of internal symmetries is closely related to
particle/field statistics\footnote{%
The analysis of statistics from first principles leads rather directly to
parastatistics in the sense of \cite{Ha}. It is one of the great
achievements of particle physics in the 80$^{ies}$ to show that this may be
always converted into Fermion/Boson statistics + internal group symmetry
where the latter can be computed from the structure of the structure of the
causal observables.}, there is some subtle problem with the Klein-Kaluza
mechanism in QFT away from the quasiclassical pictures of functional
integrals.

Another problematic point is the intrinsic meaning of ``stringyness'' in
form of an infinite tower of particles with an oscillator-like mass
spectrum. As long as mass spectra do not accumulate (by increase of
multiplicities) too densely, they are compatible with the phase space
structure of QFT and lead to reasonable thermal behavior, i.e. the
pathological situation of a finite Hagedorn-temperature can presumably also
be avoided in string theory. But it is not known to me how one can
distinguish an infinite collection of resonances i.e. poles in the second
Riemann sheet (since presumably in string theory most of the particles in
the tower are unstable through higher order (gereri) interactions as it
would be the case in Feynman theory). I do not know of any theorem in QFT
which forbids such a resonance situation and therefore I do not understand
the meaning of stringyness. Extended objects can also exist in QFT build on
perfect local observables; in fact the superselection theory even demands in
some cases the existence of noncompactly localized objects which intertwine
between inequivalent representations of perfectly local observable algebras,
examples are the carriers of braid group statistics in d=1+2 dimensions are
necessarily extended along semiinfinite spacelike strings. So it is very
questionable if there exists an intrinsic meaning of stringyness.

The relation of string theory with the wedge-localization approach to QFT
presented in this paper goes only via the common historical root of the
S-matrix theory of the 60$^{ies}$ and basically consists in the claim that
both theories are ultrviolet finite.$.$ In fact the on-shell nature of wedge
algebras as exemplified by the modular wedge-localization equation (\ref
{modloc}) provides a field theoretic link for the S-matrix bootstrap and
transports the ultraviolet finiteness of the latter into QFT. Although this
finiteness is shared with string theory, the cause of it is very different.
Wheras in string theory this finiteness results from the extension\footnote{%
This means that the string of string theory is not an extended object in an
otherwise local theory as. e.g. a Mandelstam string in gauge theory.} of a
string as an indecomposable state of matter, the modular approach to QFT is
ultraviolet-finite in a much more radical and at the same time much more
conservative way. The radical aspect is that by not using the inevitably
light-cone singular field coordinatizations in the actual construction but
rather a net of algebras, the objects to which the the bad short distance
behaviour and the ultraviolet divergencies is attached have disappeared from
the scene. They may be constructed at the end as local generators of the
already constructed spacetime-indexed nets of algebras, but there they
cannot do no harm anymore. The conservative aspect is that by taking this
approach which requires a radically changed formalism, one remains in total
harmony with the causality-, spectral-, and degrees of freedom principles
which underlie QFT. The short distance behavior of the field approach is
substituted by the nontriviality of intersections of algebras. This approach
has already been tested in the bootstrap-formfactor constructions of d=1+1
factorizable models. In d=1+3 one expects that its perturbative version
reproduces the renormalizable field theories and in addition reveals whether
the frontiers of the standard approach (renormalizable/nonrenormalizable)
which appear in a purely formal way (power counting in auxiliary objects)
are really the the intrinsic formalism-independent frontiers defined by the
physical principles of QFT. Massive gauge theories analyzed from the
slightly physical point of view \cite{Du-Sch} of selfinteracting massive
vectormesons nourish the suspicion that the intrinsic frontiers may be wider
than those set by the standard perturbative power counting for interaction
polynomials.

Both the modular wedge localization approach as well as string theory
attribute a basic significance to chiral conformal theory, and both know the
notion of holography. But the use and the physical interpretation of these
concepts is quite different. Whereas in AQFT chiral conformal theories are
the building blocks of holographic images of higher dimensional theories and
therefore are positioned in the same Minkowski space, string theory places
the chiral conformal data into an auxiliary source space and identifies the
physical space as the target space of the fields in which they take their
values. Related to this is in fact the notorious difficulty of defining a
string field theory, a problem which is presumably related to the difficulty
in separating the intrinsic conceptual content of string theory from its
procedural prescriptions.

On the other hand the modular approach has all the hallmarks of a
conceptually based intrinsic formulation of local particle physics which
makes it a candidate for an extension into the realm of interactions of the
Wigner's representation theory of free particles which was the first totally
intrinsic (independent of quantization) approach to relativistic quantum
theory.

\textbf{Note added:} Meanwhile there has been substancial progress on an
issue, which although dating back to the QFT of the 70$^{ies},$ received new
attention through the string theorist's observation about a correspondence
of \ QFT in anti De Sitter space and conformal QFT. The methods of AQFT have
turned out to be very powerful in unravelling some real time quantum
physical aspect of this isomorphism and have led to substancial progress on
higher dimensional conformal QFT (B. Schroer hep-th/0005010 and 0005134).
The role of string theory as a search machine of such unexpected connections
versus the conceptual power of AQFT to analyse and understand such
observations on a profound and useful level for particle physics makes an
interesting continuation of the theme of the last section in which the
reader may be interested in.

\begin{acknowledgement}
A correspondence with K.-H. Rehren, and discussions with J. Roberts and R.
Longo during a short visit of Rome university led to various improvements of
the paper. I am particularly indebted to J. Roberts for the invitation which
made these discussions possible. Finally I would like to thank my colleagues
at the FU Andreas Fring, Michael Karowski, Hradch Babujian and Robert
Schrader for some good questions and constructive suggestions.
\end{acknowledgement}

\appendix

\section{Appendices on Modular Theory}

\subsection{Some Facts about Modular Theory}

\begin{definition}
A von Neumann algebra $\mathcal{A}$ (weakly closed operator sub-algebra of
the full algebra B(H) on a Hilbert space H) is in ``standard'' position''
with respect to a vector $\Omega \in H,$ denoted as ($\mathcal{A},\Omega ),$
if $\Omega $ is a cyclic ($\overline{\mathcal{A}\Omega }=H)$ and separating (%
$A\Omega =0,A\in \mathcal{A}$ iff $A=0$) vector for $\mathcal{A}.$ In this
situation Tomita defines the following involutive antilinear but unbounded
operator (the Tomita involution S) 
\begin{equation}
SA\Omega :=A^{\ast }\Omega  \label{Tom}
\end{equation}
where the star operation is the hermitian conjugate in operator algebras.
Its closability property (as physicists we will use the same notation for
the closure) is the prerequisite for the polar decomposition 
\begin{equation}
S=J\Delta ^{\frac{1}{2}}
\end{equation}
where the angular part $J$ (the modular involution) is antiunitary with $%
J^{2}=1$ and $\Delta $ is unbounded positive and therefore leads to a
unitary group $\Delta ^{it}.$
\end{definition}

\begin{theorem}
(Tomita 1965, with significant improvements from Takesaki): The modular
involution maps $\mathcal{A}$ onto its von Neumann commutant $\mathcal{A}%
^{\prime }$ in $H$: 
\begin{equation}
AdJ\cdot \mathcal{A}=\mathcal{A}^{\prime }
\end{equation}
The unitary $\Delta ^{it}$ defines a ``modular'' automorphism group by 
\begin{equation}
Ad\Delta ^{it}\cdot \mathcal{A}=\mathcal{A}
\end{equation}
(analogy to a dynamical law for the algebra).
\end{theorem}

More details and references to the proof can be found in \cite{Ha}. Actually
physicists have independently discovered some important properties of
modular theory which later were incorporated by mathematicians into the
Tomita-Takesaki theory. In fact Haag, Hugenholtz and Winnink\cite{Ha}
observed that the KMS property which Kubo, Martin and Schwinger just used as
a computational trick in order to avoid the calculation of traces in quantum
statistical mechanics took on a fundamental conceptual role if one works
directly in the thermodynamic limit of infinitely extended systems. a
closely related independent discovery in their pursuit of
physical-conceptual problems in quantum statistical mechanics which arise if
one works directly in the thermodynamic limit \cite{Ha}. As it is well
known, the Gibbs representation formula 
\begin{eqnarray}
\left\langle A_{V}\right\rangle _{\beta }^{(V)} &=&\frac{tre^{-\beta
H_{V}}A_{V}}{tre^{-\beta H_{V}}} \\
A_{V} &\in &algebra\,\,\,of\,\,\ box-quantization  \notag
\end{eqnarray}
ceases to make sense\footnote{%
In a box the bounded below hamiltonian aquires a discrete spectrum and e$%
^{-\beta H}$ is of trace class ($\Omega _{\beta }=e^{-\frac{1}{2}\beta H}$
is H.S.), a property which is lost in the infinite volume limit.} for
infinite volume open systems and the algebra changes its Murray von Neumann
type. Whereas in the quantization box it was type I, the open system algebra
becomes type III$_{1}$ and the Gibbs formula passes to the KMS condition
which is a cyclic relation for thermal correlation functions \cite{Ha}. In
the 70$^{ies}$ Haag and collaborators were able to derive the KMS condition
directly from stability properties under local deformations and Pusz and
Woronowicz found a direct link to the second law of thermodynamics \cite{Ha}%
. These profound results were recently used for the derivation of thermal
properties of quantum matter in an anti-de Sitter spacetime \cite{BFS}.

The relation of modular theory with the Einstein causality of observables
and locality of fields in QFT was made around 1975 in a series of papers by
Bisognano and Wichmann \cite{Ha}. Specializing to wedge algebras $\mathcal{A}%
(W)$ generated by Wightman fields, they proved the following theorem

\begin{theorem}
The Tomita modular theory for the wedge algebra and the vacuum state vector (%
$A(W),\Omega )$ yields the following physical identifications 
\begin{eqnarray}
\Delta ^{it} &=&U(\Lambda _{W}(2\pi t)) \\
J &=&TCP\cdot U(R_{x}(\pi ))  \notag
\end{eqnarray}
\end{theorem}

Here $\Lambda _{W}(\chi )$ denotes the boost ($\chi $ is the x-space
rapidity) which leaves the wedge $W$ invariant. If we choose the standard $t$%
-$x$ wedge, then the rotation which aligns the TCP with Tomita's $J$ is a
rotation around the x-axis by an angle $\pi .$

Now I come to my own contributions which are of a more recent vintage \cite
{Sch1}. They result from the desire to invert the Bisognano-Wichmann theorem
i.e. to use Tomita's modular theory for the actual construction (and
classification) of (a net of) wedge algebras belonging to interacting
theories with the final goal to intersect wedge algebras in order to obtain
a net of compactly localized double cone algebras. For the arguments which
show that the particle physics properties, in particular the scattering
matrix and formfactors of distinguished fields (conserved currents) can be
abstracted from the net observables, I refer to \cite{Ha}\cite{Bu}\cite{B-H}%
. If desired. the nets can also be coordinatized by more traditional
pointlike fields and a rigorous derivation for chiral nets can be found in 
\cite{Fr-Jo}. For the derivation of LSZ scattering theory one makes the
assumption of the existence of a mass gap. With this one immediately
realizes that, whereas the connected part of the Poincar\'{e} group is the
same as that of the free incoming theory, the disconnected part containing
time reversals, in particular the modular involution $J$ for the wedge carry
the full interaction 
\begin{eqnarray}
\Delta _{W}^{it} &=&\Delta _{W,in}^{it}=:e^{-iKt}  \label{rel} \\
J_{W} &=&S_{sc}J_{W,in}  \notag
\end{eqnarray}
Here $J_{W,in}$ refers to the Tomita involution (or TCP reflection) of the
wedge algebra generated by the incoming free field. If the theory is not
asymptotically complete (i.e. the vacuum is not cyclic with respect to the
incoming fields) these relations have to be modified, but here we discard
such pathologies for which no physical illustration exists. Since we do not
want to temper with historically grown notations, we have added a subscript
to the S-matrix $S_{sc}$ in order to distinguish it where necessary from
Tomita's $S.$ The modular ``Hamiltonian'' $K$ defined in the first equation
(the boost generator= Hamiltonian of a particular uniformly accelerated
Unruh observer) has always symmetric instead of one-sided spectrum.

The last relation (\ref{rel}) is nothing but the TCP-transformation law of
the S-matrix rewritten in terms of modular objects associated to the wedge
algebra. The above role of the S-matrix as a kind of relative modular
invariant of the wedge algebra (relative to the free one) is totally
characteristic for \textit{local} quantum physics and has no counterpart in
quantum mechanics.

\subsection{Absence of PFGs for Sub-Wedge Regions in Theories with
Interactions}

\begin{theorem}
In interacting theories there exist no PFGs localized in subwedge regions.
The wedge region is the smallest spacetime region for which PFGs in the
presence of interactions are possible.
\end{theorem}

For the proof\footnote{%
The proof is similar to that of the Jost-Schroer theorem in \cite{St-Wi} and
to that in \cite{Mund}.} let us first assume that the spacetime localization
region $\mathcal{O}$ of the would be PFGs is compact, e.g. a double cone.
Let $\phi $ be an operator which is affiliated with $\mathcal{A}(\mathcal{O}%
) $ which means that on the domain of definition it commutes with all
operators from the commutant $\mathcal{A}^{\prime }(\mathcal{O}).$ The PFG
property of $\phi $ means ($\phi ^{\#}$ stands for either $\phi $ or $\phi
^{\ast }$) 
\begin{eqnarray}
\phi ^{\#}(x)\Omega &=&one-particle\,\,vector \\
\phi ^{\#}(x) &=&U(x)\phi ^{\#}U^{\ast }(x)  \notag
\end{eqnarray}
without any admicture of additional polarization contribution from higher
particle configurations. As a result the vector satisfies the free field
equation in x. On the other hand we have that $\left[ \phi ^{\#}(x),\phi
^{\#}(y)\right] =0$ for sufficiently large spacelike separations. Let us now
look at matrix-elements 
\begin{equation}
\left\langle \psi _{2}\left| \phi ^{\#}(x)\right| \psi _{1}\right\rangle
\end{equation}
with say $\psi _{2}\in $ domain($\phi ^{\#}$) and choose $\psi _{1}$ from
the dense set of state vectors which are localized in some region spacelike
relative to loc($\phi ^{\#}(x)).$ This is done by applying spacelike
separated operators onto the vacuum $\left| \psi _{1}\right\rangle =A\left|
0\right\rangle $. Since $\phi ^{\#}(x)$ commute with such operators we
obtain 
\begin{eqnarray}
\left\langle \psi _{2}\left| \phi ^{\#}(x)\right| \psi _{1}\right\rangle
&=&\left\langle \psi _{2}\left| \phi ^{\#}(x)A\right| \Omega \right\rangle \\
&=&\left\langle \psi _{2}\left| A\phi ^{\#}(x)\right| \Omega \right\rangle 
\notag
\end{eqnarray}
i.e. $\phi ^{\#}(x)$ fulfills the free field equation on a dense set of
states in its domain. Since all affiliated operators are closable, the
operator itself fulfills the free field equation. If we succeed to prove in
addition that the commutator with itself is a c-number 
\begin{equation}
\left[ \phi ^{\ast }(x),\phi (y)\right] =c(x-y)\mathbf{1}
\end{equation}
then we would have achieved our goal since it would follow that $\phi
^{\#}(x)$ is a linear expression in terms of the particle creation and
annihilation operator which contradicts the presence of an interaction. But
this last step follows almost literally the argument in the derivation of
the Jost-Schroer theorem \cite{St-Wi}, the fact that the present $\phi $ has
no well defines L-covariance does not matter. In the first step one shows
that 
\begin{equation}
\left[ \phi ^{\ast }(x),\phi (y)\right] \left| \Omega \right\rangle
=c(x-y)\left| \Omega \right\rangle
\end{equation}
which requires the creation$\times $creation contribution to vanish i.e. $%
\left[ \phi ^{\ast (+)}(x),\phi ^{(+)}(y)\right] \left| \Omega \right\rangle
=0.$ For this one uses causality and the separate analyticity in x and y
which follows from the forward mass-shell support property. The
generalization from a relation on the vacuum to a relation on a dense set of
states is as before.

\end{document}